\PassOptionsToPackage{prologue,table}{xcolor}
\documentclass[manuscript,screen,review=false]{acmart}

\usepackage{multirow}
\usepackage{threeparttable}
\usepackage{amsmath}
\usepackage{bm}
\usepackage{amssymb}
\usepackage{tabularx}
\usepackage{multirow}
\usepackage{alphabeta}
\usepackage{amsthm}
\usepackage{amsmath}
\usepackage{booktabs} 
\usepackage{threeparttable}
\usepackage{bbding}
\usepackage{pifont}
\usepackage{wasysym}
\usepackage{fontawesome}
\usepackage{svg}
\usepackage{makecell}
\usepackage{amsmath,amsfonts}
\usepackage{algorithmic}
\usepackage{algorithm}
\usepackage{array}
\usepackage{soul}
\usepackage{fixltx2e}

\AtBeginDocument{%
  }

\setcopyright{acmlicensed}
\copyrightyear{2024}
\acmYear{2024}
\acmDOI{XXXXXXX.XXXXXXX}

\acmBooktitle{Woodstock '18: ACM Symposium on Neural Gaze Detection,
  June 03--05, 2018, Woodstock, NY}
\acmISBN{978-1-4503-XXXX-X/18/06}

\begin{document}

\title{Intellectual Property Protection for Deep Learning Model and Dataset Intelligence}

\author{Yongqi Jiang}
\email{jiangyongqi@njust.edu.cn}
\affiliation{%
 \institution{Nanjing University of Science and Technology}
 \city{Nanjing}
 \country{PR China,}}
 \affiliation{%
  \institution{State Key Laboratory of Integrated Services Networks (Xidian University)}
  \city{Xi'an}
  \country{PR China}}
 
\author{Yansong Gao}
\email{garrison.gao@uwa.edu.au}
\affiliation{%
 \institution{The University of Western Australia}
 \city{Perth}
 \country{Australia}}

\author{Chunyi Zhou}
\email{zhouchunyi@zju.edu.cn}
\affiliation{%
 \institution{Zhejiang University}
 \city{Hangzhou}
 \country{PR China}}

\author{Hongsheng Hu}
\email{Hongsheng.Hu@newcastle.edu.au}
\affiliation{%
 \institution{University of Newcastle}
 \city{Newcastle}
 \country{Australia}}

\author{Anmin Fu\textsuperscript{*}}
\email{fuam@njust.edu.cn}
\affiliation{%
 \institution{Nanjing University of Science and Technology}
 \city{Nanjing}
 \country{PR China,}}
\affiliation{%
 \institution{State Key Laboratory of Integrated Services Networks (Xidian University)}
 \city{Xi'an}
 \country{PR China}}

\author{Willy Susilo}
\email{wsusilo@uow.edu.au}
\affiliation{%
 \institution{University of Wollongong}
 \city{Wollongong}
 \country{Australia}}

\renewcommand{\thefootnote}{\fnsymbol{footnote}}
\footnotetext[1]{Corresponding authors.}

\renewcommand{\shortauthors}{Jiang et al.}

\begin{abstract}
\par With the growing applications of Deep Learning (DL), especially recent spectacular achievements of Large Language Models (LLMs) such as ChatGPT and LLaMA, the commercial significance of these remarkable models has soared. However, acquiring well-trained models is costly and resource-intensive. It requires a considerable high-quality dataset, substantial investment in dedicated architecture design, expensive computational resources, and efforts to develop technical expertise. 
Consequently, safeguarding the Intellectual Property (IP) of well-trained models is attracting increasing attention. In contrast to existing surveys overwhelmingly focusing on model IPP mainly, this survey not only encompasses the protection on model level intelligence but also valuable dataset intelligence. Firstly, according to the requirements for effective IPP design, this work systematically summarizes the general and scheme-specific performance evaluation metrics. Secondly, from proactive IP infringement prevention and reactive IP ownership verification perspectives, it comprehensively investigates and analyzes the existing IPP methods for both dataset and model intelligence. Additionally, from the standpoint of training settings, it delves into the unique challenges that distributed settings pose to IPP compared to centralized settings. Furthermore, this work examines various attacks faced by deep IPP techniques. Finally, we outline prospects for promising future directions that may act as a guide for innovative research.
\end{abstract}

\maketitle

\section{Introduction}

Deep learning (DL) has revolutionized a wide variety of artificial intelligence (AI) applications, including the prediction of 3D structures for proteins \cite{jumper2021highly}, self-driving cars \cite{ndikumana2020deep}, and the relatively new artificial intelligence-generated content (AIGC) \cite{lawrence2024ecological}. Despite these surprising advances, it is widely understood that building a production-level DL model is a non-trivial task, which frequently calls for (i) meticulously designed network structures, (ii) a large-scale, high-quality (annotated) dataset, (iii) a massive amount of computational resources, and (iv) high technical expertise. It is therefore extremely costly for acquisition of high-performance DL models, especially recent large models like ChatGPT \cite{Walker23}, diffusion models (DMs) \cite{fernandez2023stable}, Codex \cite{lawrence2024ecological}, etc. These well-trained models and their associated training datasets represent valuable intellectual property (IP) and form the core competencies of companies.

Unfortunately, the immense value of such IP incentivizes adversaries to launch various model or dataset theft attacks. For instance, the work \cite{lv2024mea} mimics the original model at a meager cost by querying and observing the outputs. 
The unauthorized reproduction and malicious distribution of well-trained models result in copyright infringement and significant financial losses for the model creators.
Besides, with the auxiliary of prior arts like member inference, preference profiling, model inversion attacks, etc. \cite{Zhou23}, attackers can reconstruct the model’s private training data. Recently, the concept of ``Machine Learning as a Service (MLaaS)” has gained popularity, allowing individuals who are not DL experts to remotely access product-level models on a pay-per-use basis. However, MLaaS also introduces new avenues for attackers, who can frequently request prediction results from these services to create low-cost, functionally identical stolen models. This ability enables attackers to avoid payment, create services that violate IP rights, and potentially gain significant financial profits, thereby undermining the competitive edge of the original IP owners in the market.

Deep IP protection (IPP) is a frontier research field that is still in its infancy. Researchers have proposed various IPP methods~\cite{nie2024deep,liu2022your,tian2023knowledge}, most of which fall under the category of reactive verification, where IP identifiers (watermarks or fingerprints) are embedded into models and then used to verify the ownership of suspicious models. The advantage of watermark is that it can provide more accurate ownership verification and even maintain a certain degree of robustness in transfer learning scenarios \cite{tian2023knowledge}. However, these techniques are invasive; they require tampering with the training process, which may affect model utility or introduce new security risks. In contrast, fingerprinting can better preserve the model's functionality. It also demonstrates more versatility and convenience, especially in scenarios where model owners do not have the right to modify the model \cite{liu2022your}. However, it also has limitations when facing diverse and adaptive attacks. The other deep IPP category is the proactive approach to preventing model theft~\cite{xue2023dataset,ren2022protecting,lv2023robustness}. Controlling model access and tracking leakers of pirated models are components of reactive protection strategies. Common measures include access authentication \cite{xue2023dataset} or entangling the model usability and access privilege \cite{ren2022protecting}. Nonetheless, deep IPP regardless of whether reactive or proactive faces a range of attack types, where the capabilities of attackers can vary. 

\par Designing an ideal IPP scheme is thus challenging. On the one hand, trade-offs among multiple competitive objectives have to be balanced, such as ensuring the watermark's invisibility while also possessing robustness. 
On the other hand, it is imperative to guarantee that the IPP scheme functions properly in various application circumstances. For example, if a DL model is trained and watermarked on a skin disease dataset, and the service buyer fine-tunes the model on a pneumonia dataset, the IPP should survive the ownership verification procedure. 

\subsection{Contributions of this survey}

\par Deep IPP is currently in its early stages, with a notable lack of comprehensive summaries and analyses to better characterize the state-of-the-art. 
Therefore, conducting a systematic and comprehensive survey on the current status and advancements in this field is crucial. To this end, this review spans key research from 2017 to 2024, focusing on copyright protection for both models and datasets (the latter often being overlooked). It covers the taxonomy of IPP algorithms under centralized and decentralized learning settings, evaluation metrics, and various attacks that pose threats to IPP algorithms. Furthermore, we identify core issues and major challenges in IPP, outline promising future research directions, and highlight practical applications. Our major contributions can be summarized as follows:
\begin{itemize}
\item We conduct a comprehensive review of existing deep IPP schemes for both DL models and datasets with particular attention to the unique challenges and solutions associated with AIGC models such as DMs and LLMs (Section \ref{3.2} and \ref{3.3}). It fills overlooked aspects in existing surveys, which overlook the fact that datasets are also important and valuable intellectual property.

\item We innovatively summarize dual-level performance evaluation metrics: common metrics applicable to all IPP approaches, and unique metrics specific to each type according to their distinctive defense objectives.

\item We undertake a systematic review of deep IPP and attack methods for diverse tasks. These methods are categorized from reactive and proactive defense perspectives, and a critical analysis of their strengths and limitations within each (sub)category is offered.

\item We provide an in-depth analysis of the challenges encountered by deep IPPs in distributed settings, categorizing existing methods and comparing their respective advantages and disadvantages.

\item We identify the limitations associated with deep IPPs, outlining prospect promising avenues for future research.
\end{itemize}

\subsection{Comparison with Existing Surveys}
\begin{table*}[htp]
\centering
\caption{Comparison with existing surveys.}
\label{table:1}
\renewcommand{\arraystretch}{1.5}
\scalebox{0.7}{
\begin{threeparttable}
\begin{tabular}{|c|c|cc|cc|cc|cc|cc|cc|}
\hline
\textbf{Paper} & \textbf{Year} & \multicolumn{2}{c|}{\textbf{Evaluation Metric}} & \multicolumn{2}{c|}{\textbf{Taxonomy}} & \multicolumn{2}{c|}{\textbf{Distributed IPPs}} & \multicolumn{2}{c|}{\textbf{Attacks on Deep IPPs}} & \multicolumn{2}{c|}{\textbf{Challenges and Prospects}}\\ \cline{3-8} \cline{9-12}

& & \makebox[0.05\textwidth][r]{Rea.\tnote{1}} & \makebox[0.05\textwidth][r]{Proa.\tnote{2}} & \makebox[0.05\textwidth][r]{Mod.\tnote{3}} & \makebox[0.05\textwidth][r]{Data.\tnote{4}} & \makebox[0.05\textwidth][r]{Chall.\tnote{5}} & \makebox[0.05\textwidth][r]{Fed IPP\tnote{6}} & \makebox[0.05\textwidth][r]{D\&E\tnote{7}} & \makebox[0.05\textwidth][r]{Remo.\tnote{8}} & \makebox[0.05\textwidth][r]{Chall.\tnote{5}} & \makebox[0.05\textwidth][r]{Pros.\tnote{9}} \\\hline

Xue \textit{et al.} \cite{xue2021intellectual} & 2021 & \makebox[0.04\textwidth][r]{\Checkmark} & \makebox[0.04\textwidth][r]{\XSolidBrush} 
& \makebox[0.04\textwidth][r]{\Checkmark} & \makebox[0.04\textwidth][r]{\XSolidBrush} 
& \makebox[0.04\textwidth][r]{\XSolidBrush} & \makebox[0.04\textwidth][r]{\Checkmark}  
& \makebox[0.04\textwidth][r]{\Checkmark} & \makebox[0.04\textwidth][r]{\Checkmark}
& \makebox[0.04\textwidth][r]{\XSolidBrush} & \makebox[0.04\textwidth][r]{\Checkmark} \\ \hline

Li \textit{et al.} \cite{li2021survey} & 2021 & \makebox[0.04\textwidth][r]{\Checkmark} & \makebox[0.04\textwidth][r]{\XSolidBrush} 
& \makebox[0.04\textwidth][r]{\Checkmark} & \makebox[0.04\textwidth][r]{\XSolidBrush} 
& \makebox[0.04\textwidth][r]{\XSolidBrush} & \makebox[0.04\textwidth][r]{\XSolidBrush}  
& \makebox[0.04\textwidth][r]{\XSolidBrush} & \makebox[0.04\textwidth][r]{\XSolidBrush}
& \makebox[0.04\textwidth][r]{\XSolidBrush} & \makebox[0.04\textwidth][r]{\XSolidBrush} \\ \hline

Regazzoni \textit{et al.} \cite{regazzoni2021protecting} & 2021 & \makebox[0.04\textwidth][r]{\XSolidBrush} & \makebox[0.04\textwidth][r]{\XSolidBrush} 
& \makebox[0.04\textwidth][r]{\Checkmark} & \makebox[0.04\textwidth][r] {\XSolidBrush}
& \makebox[0.04\textwidth][r]{\XSolidBrush} & \makebox[0.04\textwidth][r]{\XSolidBrush}  
& \makebox[0.04\textwidth][r]{\Checkmark} & \makebox[0.04\textwidth][r]{\Checkmark}
& \makebox[0.04\textwidth][r]{\Checkmark} & \makebox[0.04\textwidth][r]{\XSolidBrush} \\ \hline

Lounici \textit{et al.} \cite{lounici2021yes} & 2021 & \makebox[0.04\textwidth][r]{\Checkmark} & \makebox[0.04\textwidth][r]{\XSolidBrush} 
& \makebox[0.04\textwidth][r]{\Checkmark} & \makebox[0.04\textwidth][r] {\XSolidBrush}
& \makebox[0.04\textwidth][r]{\XSolidBrush} & \makebox[0.04\textwidth][r]{\XSolidBrush}  
& \makebox[0.04\textwidth][r]{\Checkmark} & \makebox[0.04\textwidth][r]{\Checkmark}
& \makebox[0.04\textwidth][r]{\Checkmark} & \makebox[0.04\textwidth][r]{\XSolidBrush} \\ \hline

Fkirin \textit{et al.} \cite{fkirin2022copyright} & 2022 & \makebox[0.04\textwidth][r]{\XSolidBrush} & \makebox[0.04\textwidth][r]{\XSolidBrush} 
& \makebox[0.04\textwidth][r]{\Checkmark} & \makebox[0.04\textwidth][r] {\XSolidBrush}
& \makebox[0.04\textwidth][r]{\XSolidBrush} & \makebox[0.04\textwidth][r]{\XSolidBrush}  
& \makebox[0.04\textwidth][r]{\XSolidBrush} & \makebox[0.04\textwidth][r]{\XSolidBrush}
& \makebox[0.04\textwidth][r]{\Checkmark} & \makebox[0.04\textwidth][r]{\XSolidBrush} \\ \hline

Xue \textit{et al.} \cite{xue2023turn} & 2023 & \makebox[0.04\textwidth][r]{\XSolidBrush}  & \makebox[0.04\textwidth][r]{\Checkmark} 
& \makebox[0.04\textwidth][r]{\Checkmark} & \makebox[0.04\textwidth][r]{\XSolidBrush} 
& \makebox[0.04\textwidth][r]{\XSolidBrush} & \makebox[0.04\textwidth][r]{\XSolidBrush}  
& \makebox[0.04\textwidth][r]{\XSolidBrush} & \makebox[0.04\textwidth][r]{\Checkmark}
& \makebox[0.04\textwidth][r]{\Checkmark} & \makebox[0.04\textwidth][r]{\XSolidBrush} \\ \hline

Peng \textit{et al.} \cite{peng2023intellectual} & 2023 & \makebox[0.04\textwidth][r]{\Checkmark} & \makebox[0.04\textwidth][r]{\XSolidBrush} 
& \makebox[0.04\textwidth][r]{\Checkmark} & \makebox[0.04\textwidth][r]{\XSolidBrush} 
& \makebox[0.04\textwidth][r]{\Checkmark} & \makebox[0.04\textwidth][r]{\XSolidBrush}  
& \makebox[0.04\textwidth][r]{\Checkmark} & \makebox[0.04\textwidth][r]{\XSolidBrush}
& \makebox[0.04\textwidth][r]{\Checkmark} & \makebox[0.04\textwidth][r]{\Checkmark} \\ \hline

Lansari \textit{et al.} \cite{lansari2023federated} & 2023 & \makebox[0.04\textwidth][r]{\Checkmark} & \makebox[0.04\textwidth][r]{\XSolidBrush} 
& \makebox[0.04\textwidth][r]{\Checkmark} & \makebox[0.04\textwidth][r]{\XSolidBrush} 
& \makebox[0.04\textwidth][r]{\Checkmark} & \makebox[0.04\textwidth][r]{\Checkmark}  
& \makebox[0.04\textwidth][r]{\Checkmark} & \makebox[0.04\textwidth][r]{\Checkmark}
& \makebox[0.04\textwidth][r]{\XSolidBrush} & \makebox[0.04\textwidth][r]{\Checkmark} \\ \hline

Lederer \textit{et al.} \cite{lederer2023identifying} & 2023 & \makebox[0.04\textwidth][r]{\Checkmark} & \makebox[0.04\textwidth][r]{\XSolidBrush} 
& \makebox[0.04\textwidth][r]{\Checkmark} & \makebox[0.04\textwidth][r] {\XSolidBrush}
& \makebox[0.04\textwidth][r]{\XSolidBrush} & \makebox[0.04\textwidth][r]{\XSolidBrush}  
& \makebox[0.04\textwidth][r]{\Checkmark} & \makebox[0.04\textwidth][r]{\Checkmark}
& \makebox[0.04\textwidth][r]{\Checkmark} & \makebox[0.04\textwidth][r]{\XSolidBrush} \\ \hline

 Sun \textit{et al.} \cite{sun4697020deep} & 2024 & \makebox[0.04\textwidth][r]{\Checkmark} & \makebox[0.04\textwidth][r]{\XSolidBrush} 
& \makebox[0.04\textwidth][r]{\Checkmark} & \makebox[0.04\textwidth][r]{\XSolidBrush} 
& \makebox[0.04\textwidth][r]{\XSolidBrush} & \makebox[0.04\textwidth][r]{\Checkmark} 
& \makebox[0.04\textwidth][r]{\Checkmark} & \makebox[0.04\textwidth][r]{\Checkmark} 
& \makebox[0.04\textwidth][r]{\XSolidBrush} & \makebox[0.04\textwidth][r]{\Checkmark} \\ \hline

\rowcolor{blue!5}\textbf{Ours} & \textbf{2024} & \makebox[0.04\textwidth][r]{\Checkmark} & \makebox[0.04\textwidth][r]{\Checkmark} 
& \makebox[0.04\textwidth][r]{\Checkmark} & \makebox[0.04\textwidth][r] {\Checkmark}
& \makebox[0.04\textwidth][r]{\Checkmark} & \makebox[0.04\textwidth][r]{\Checkmark}  
& \makebox[0.04\textwidth][r]{\Checkmark} & \makebox[0.04\textwidth][r]{\Checkmark} 
& \makebox[0.04\textwidth][r]{\Checkmark} & \makebox[0.04\textwidth][r]{\Checkmark} \\ \hline
\end{tabular}
\begin{tablenotes}
\footnotesize
\item(1) Rea. -- Reactive 
\quad(2) Proa. -- Proactive  
\quad(3) Mod. -- Model IPP  
\quad(4) Data. -- Dataset IPP  
\quad(5) Chall. -- Challenges
\quad(6) Fed IPP -- Federated Learning IPP 
\quad(7) D\&E -- IP Detection \& Evasion
\quad(8) Remo. -- IP Removal
\quad(9) Pros. -- Prospects 
\centering
\item {\Checkmark} indicates that the proposed survey has considered this aspect, while {\XSolidBrush} indicates that it has not.
\end{tablenotes}
\end{threeparttable}}
\end{table*}

\par  Our survey advances the examination of IPP on not only model and but also dataset intelligence from a multitude of crucial aspects, as summarized in Table \ref{table:1}. Firstly, we systematically identify the common performance evaluation metric for both reactive and proactive IPP categories and further differentiate the unique metric specific to each given their distinctive defense objectives. 
This aims to address the inadvertent limitations of existing surveys, which either generalize or only discuss the performance evaluation metric for reactive IPP, providing a more comprehensive and refined evaluation framework. 
Secondly, upon taxonomy, we elaborate on reviewing methods of protecting the IP of DL models and datasets. The latter is often missing in existing surveys. 
Thirdly, we delve into the challenges of IPP in distributed learning, examining existing IPP methods, which are underexplored in existing surveys.
Fourthly, we classify various types of threats that IPP faces and thoroughly analyze the goal of the adversary and the underlying assumptions. It is noted that most existing surveys often focus solely on threats to reactive IPP methods, while ignoring the evolving new attacks targeting proactive IPP schemes, thus limiting the scope of countermeasures and foresight of the survey to some extent. 
Finally, we draw attention to the challenges associated with IPP and prospect future directions.

\subsection{Organization}
\par The rest of this work is organized as follows. Section \ref{2} presents the background of deep IPP, including the definition of deep neural networks, deployment and verification modes, and evaluation metrics. In Sections \ref{3}, \ref{3.1} and \ref{4}, we categorize the deep IPP for mode and dataset intelligence, respectively. Section \ref{5} investigates the emerging deep IPP in distributed learning settings. Security threats to deep IPs are investigated in Section \ref{7}. Following, we delineate the current challenges facing deep IPP and propose auspicious directions for prospective research in Section \ref{8}. Section \ref{9} concludes this survey.

\section{Background}
\label{2}
\subsection{Deep Neural Networks}
\par A DL model is made up of a number of network layers, including an input layer, several hidden layers, an output layer, and so on. The DL model maps input data to its corresponding label (using classification as an example) by employing the approximate transformation function $\Phi$ :
\begin{equation}\label{eq.1}
\mathop {\min }\limits_\theta  \sum\limits_{i = 1}^N {\frac{1}{N}{\mathcal{L}_{ce}}(\Phi ({x_i};\theta ),{y_i})},
\end{equation}  
where $N$ is the number of samples in the training dataset $\mathcal{D} = \left\{ {({x_i},{y_i})} \right\}_{i = 1}^N$; $\mathcal{L}_{ce}$ is the cross entropy (CE) loss function.
\par DL models learn the data representation through multi-layer nonlinear transformations, and optimize the model by adjusting the trainable weights so that it can accurately perform classification, prediction, or other tasks.

\subsection{Intellectual Property in MLaaS}
\par MLaaS refers to cloud-based services provided by companies for deploying machine learning (ML) products, which are available for developers and external users.
MLaaS offers two main product services: high-quality datasets and well-performed models.
However, some participants may attempt to steal these products by masquerading as customers and then rebranding and reselling them to gain illegal profits.
The stealing process is usually costless compared with obtaining a well-trained model from scratch.

\par MLaaS services are operated in two primary ways \cite{sun4697020deep,peng2023intellectual}: (1) Companies directly distribute products to buyers and grant them full access to both model and dataset, encompassing the model's internal structure and parameters as well as data properties. This mode allows users to gain insights into the model’s functionality and conduct thorough verification and review. It can be considered a white box deployment.
(2) Alternatively, companies deploy the model and dataset on their cloud server and restrict users from purchasing query rights for products without access to the product internals. Users can query task samples through released APIs to obtain results from the model output, known as black box mode.
\par Both modes have their pros and cons. Consumers are likely favoring white box mode, but the model and dataset face a high risk of leakage to the service provider, such as susceptibility to direct copy and fine-tuning attacks. In contrast, black box mode offers better protection for the confidentiality of the model and dataset, but attackers can still create function-similar pirated models by observing the model’s predictions.

\subsection{Evaluation Metrics}
\par This survey summarizes the evaluation metrics for deep IPP into common metrics and specific metrics for proactive and reactive IPP, respectively. These metrics are applicable for both model and dataset IPP evaluations.

The common metrics should be able to evaluate the following four performance measures:
\begin{itemize}
\item \textit{Robustness}: To fit downstream tasks, it is frequently necessary to make additional modifications to the protected model and the model trained on the protected dataset \cite{lv2023robustness}. Malicious attackers commonly endeavor to circumvent ownership verification through operations such as watermark deletion, overwriting, or sabotage. Thus, an ideal protection scheme should withstand diverse attacks.
\item{\textit{Efficiency}: Additional resource costs associated with the reactive verification or proactive defense approach \cite{chen2021stealing}, such as latency, and communication overhead should be affordable by the user.}
\item{\textit{Secrecy}: IP identifiers must remain secret or undetectable to adversaries \cite{lao2022identification}. This requires embedding these identifiers in a way that they are imperceptible during normal model operation and scrutiny by potential attackers, thereby safeguarding the IP from unauthorized access, reverse engineering, and tampering.}
\item{\textit{Generality}: The ideal IPPs should be agnostic to model architecture and downstream tasks \cite{li2023plmmark}.}
\end{itemize}

\par In addition to the above common metrics, a well-designed reactive ownership verification method should fulfill the following three properties:
\begin{itemize}
\item \textit{Fidelity}\textsubscript{Rea.}: IP identifiers for datasets and models are typically created by adjusting the model’s parameters or decision-making behaviors, which can often negatively impact the original model \cite{jia2022subnetwork}. Fidelity ensures that the protected model performs indistinguishably to the original model.

\item \textit{Capacity}: It refers to the valid information payload and theoretical upper bound contained in the reactive IP identifier \cite{guo2018watermarking}. An ideal IPP method must be capable of embedding a substantial amount of information in the protected DL model.

\item{\textit{Reliability}\textsubscript{Rea.}: The reactive approach should exhibit a low false negative rate \cite{lv2023robustness}, ensuring IP owners can accurately identify their IP identifiers with high confidence from the suspect models.}

\end{itemize}
\par The metrics for proactive IPP, such as efficiency, robustness, generalization, and secrecy, are identical to the reactive IPP. Moreover, fidelity and reliability deviate from the mentioned reactive metric, and there are also unique metrics for proactive methods, as outlined below.
\begin{itemize}
\item \textit{Fidelity}\textsubscript{Proa.}: The proactive authorization method adjusts fidelity according to user types. Authorized users gain superior model performance, while unauthorized users experience restricted or poor performance \cite{chen2018protect}.

\item \textit{Reliability}\textsubscript{Proa.}: To prevent theft from illegal users, it is necessary to distinguish the identity of legitimate users and illegal users accurately \cite{luo2021hierarchical}.

\item \textit{Scalability}: An ideal proactive authorization method should generate and accommodate a vast array of user identity keys~\cite{xue2023turn}.

\item \textit{Uniqueness}: The identity keys, which are allocated to legitimate users one by one, must be unique for tracking down the traitors \cite{chen2018protect}.

\item \textit{Unforgeability}: The identity key must be unforgeable, and the user identity forged by the attacker cannot be authenticated \cite{xue2023turn}.

\item \textit{Traceability}: The victim can trace out the traitor users based on the suspicious models \cite{xue2023turn}.
\end{itemize}

\section{Reactive Model Intellectual Property Protection}
\label{3}
\subsection{Overview}

We first distinguish between two types of IPP methods: (i) reactive IPP strategies, which respond to threat events after the infringement has already happened such as through verifying an implanted identifier to claim the ownership, and (ii) proactive IPP strategies, where defenders take proactive actions to stop threats before they happen such as through locking the model and later authorizing the usage right. 

\par
\begin{figure*}[htbp]
	\centering
	\includegraphics[width=.85\textwidth]{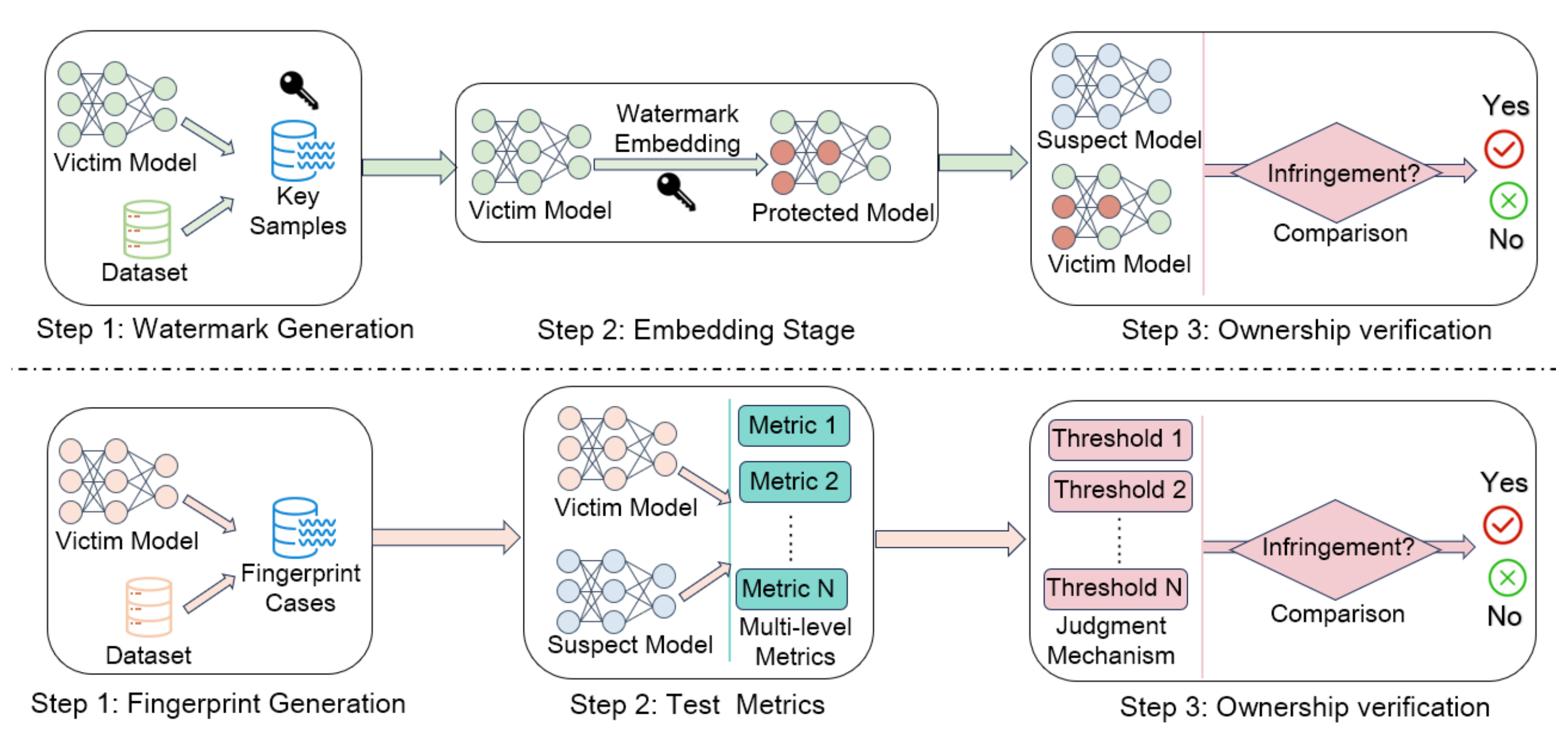}
 \vspace{-0.1in}
    \caption{The pipeline of (Top) watermarks and (Bottom) fingerprints.}
    \label{fig:WM and FP pipeline}
 \vspace{-0.1in}
\end{figure*}

 According to whether the target model’s internal components are invasively altered, reactive model IPP techniques fall into two primary subcategories: model watermarking \cite{uchida2017embedding} and model fingerprinting \cite{chen2019deepattest}. The former is invasive and the latter is non-invasive. The pipeline of model watermarking and fingerprinting is illustrated in Fig. \ref{fig:WM and FP pipeline}.

\begin{itemize}
\item{\textit{Model watermarking}: It is a solution that invasively embeds detectable and tamper-proof watermarks into the host DL model through methods such as a parameter regularizer or backdoor data watermarking. Later, the watermarks can be extracted from the model parameters or output as the ownership evidence.}
\item{\textit{Model fingerprinting}: DL fingerprinting represents a non-invasive solution wherein the model owner generates unique sample pairs that can exclusively and accurately fingerprint the target model. During the verification phase, defenders extract fingerprint samples to compare model fingerprint similarity. These disparities may manifest in various model features, such as model predictions, decision boundaries (DBs), etc.}
\end{itemize}

\subsection{Model Watermarking}
\label{3.2}
\par The existing schemes are classified according to the defender’s access level to the suspect model during verification. The comparison of these three categories is shown in Table \ref{table:model WM comparison}. 

\begin{table*}
\centering
\caption{Comparison of model watermarking approaches.}
\label{table:model WM comparison}
\renewcommand{\arraystretch}{1.2}
\scalebox{0.7}{
\begin{tabular}{|cl|c|c|l|l|}
\hline
\multicolumn{2}{|c|}{\textbf{Category}}                                                                                                                                                                                                                                                                                                           & \textbf{\begin{tabular}[c]{@{}c@{}}Model\\ Access\end{tabular}}                & \textbf{\begin{tabular}[c]{@{}c@{}}Dataset\\ Access\end{tabular}}                     & \multicolumn{1}{|c|}{\textbf{Advantages}}                                                                                                            & \multicolumn{1}{|c|}{\textbf{Drawbacks}}                                                                                                                                         \\ \hline
\multicolumn{1}{|c|}{\multirow{3}{*}{\begin{tabular}[c]{@{}c@{}}Model \\ Component-based\\ Watermarking\end{tabular}}}     & \begin{tabular}[c]{@{}l@{}}Model weights\\ \cite{uchida2017embedding,wang2021riga,lv2023robustness,liu2021watermarking,namba2019robust}\end{tabular}                                                                 & \multirow{3}{*}{White box}                                                     & \multirow{3}{*}{Entire dataset}                                                       & \multirow{2}{*}{$\bullet$ Maintain model fidelity}                                                                              & \multirow{2}{*}{$\bullet$ Relatively complex calculation}                                                                                                  \\ \cline{2-2}
\multicolumn{1}{|c|}{}                                                                                                     & \begin{tabular}[c]{@{}l@{}}Dynamic parameters\\ \cite{bharne2024enhancing,rouhani2019deepsigns,li2022move,lim2022protect}\end{tabular}                                                                          &                                                                                &                                                                                       &                                                                                                                                 &                                                                                                                                                            \\ \cline{2-2} \cline{5-6} 
\multicolumn{1}{|c|}{}                                                                                                     & \begin{tabular}[c]{@{}l@{}}Model structures\\ \cite{xiaoxuan2021meets,zhao2021structural,chen2021you}\end{tabular}                                                                                                   &                                                                                &                                                                                       & $\bullet$ No additional model modifications                                                                                     & \begin{tabular}[c]{@{}l@{}}$\bullet$ Lack of generalization and flexibility \\ $\bullet$ Low robustness against fine-tuning\\ or compression\end{tabular} \\ \hline
\multicolumn{1}{|c|}{\multirow{4}{*}{\begin{tabular}[c]{@{}c@{}}Query-based \\ Watermarking\end{tabular}}}                 & \begin{tabular}[c]{@{}l@{}}Out-of-distribution samples\\ \cite{adi2018turning,rouhani2019deepsigns,jia2021entangled,ma2024transfer,li2024comprehensive,jia2022subnetwork}\end{tabular}                               & \multirow{4}{*}{\begin{tabular}[c]{@{}c@{}}White \& \\ Black box\end{tabular}} & \begin{tabular}[c]{@{}c@{}}Dataset \\ distribution\end{tabular}                       & \begin{tabular}[c]{@{}l@{}}$\bullet$ Key samples are easily obtained \\ $\bullet$ Possess high robustness\end{tabular}          & $\bullet$ Easily detected and removed                                                                                                                      \\ \cline{2-2} \cline{4-6} 
\multicolumn{1}{|c|}{}                                                                                                     & \begin{tabular}[c]{@{}l@{}}Samples near to decision\\ boundary\\ \cite{yang2021robust,mehta2022aime}\end{tabular}                                                                                                    &                                                                                & \multirow{3}{*}{\begin{tabular}[c]{@{}c@{}}Entire \& \\ Partial dataset\end{tabular}} & $\bullet$ High secrecy, not easily detected                                                                                     & \begin{tabular}[c]{@{}l@{}}$\bullet$ Damage the model's functionality \\ $\bullet$ Sample generation is complex\end{tabular}                                     \\ \cline{2-2} \cline{5-6} 
\multicolumn{1}{|c|}{}                                                                                                     & \begin{tabular}[c]{@{}l@{}}Natural samples attached\\ with preset trigger patterns\\ \cite{guo2018watermarking,li2019prove,nie2024deep,bansal2022certified,ren2023dimension,li2022move,szyller2021dawn}\end{tabular} &                                                                                &                                                                                       & $\bullet$ High robustness against IP removal                                                                                    & $\bullet$ Trigger pattern damages usability                                                                                                                \\ \cline{2-2} \cline{5-6} 
\multicolumn{1}{|c|}{}                                                                                                     & \begin{tabular}[c]{@{}l@{}}Synonym replacement\\ \cite{li2023protecting,he2022cater,zhao2023protecting,li2023plmmark}\end{tabular}                                                                                   &                                                                                &                                                                                       & \begin{tabular}[c]{@{}l@{}}$\bullet$ Preserve the semantics without\\ requiring significant modifications\end{tabular}          & \begin{tabular}[c]{@{}l@{}}$\bullet$ Higher computational and time cost\end{tabular}                                                                     \\ \hline
\multicolumn{1}{|c|}{\multirow{4}{*}{\begin{tabular}[c]{@{}c@{}}Generative \\ Content-based \\ Watermarking\end{tabular}}} & \begin{tabular}[c]{@{}l@{}}Watermarking auto-encoders\\ \cite{cong2022sslguard,lv2022ssl}\end{tabular}                                                                                                               & \multirow{4}{*}{Black box}                                                     & \multirow{4}{*}{Entire dataset}                                                       & \begin{tabular}[c]{@{}l@{}}$\bullet$ Effective resistance to transfer\\ learning\end{tabular}                                   & \begin{tabular}[c]{@{}l@{}}$\bullet$ Compromise the usability of\\ representations\end{tabular}                                                            \\ \cline{2-2} \cline{5-6} 
\multicolumn{1}{|c|}{}                                                                                                     & \begin{tabular}[c]{@{}l@{}}Watermarking GANs\\ \cite{zhang2021deep,wu2020watermarking, abdelnabi2021adversarial, fei2024wide,lin2024cyclegan}\end{tabular}                                                           &                                                                                &                                                                                       & \multirow{2}{*}{\begin{tabular}[c]{@{}l@{}}$\bullet$ Simple operation, easy to implement\\ $\bullet$ Low overhead\end{tabular}} & \multirow{2}{*}{\begin{tabular}[c]{@{}l@{}}$\bullet$ Sacrifice visual quality and usability\end{tabular}}                                                \\ \cline{2-2}
\multicolumn{1}{|c|}{}                                                                                                     & \begin{tabular}[c]{@{}l@{}}Watermarking DMs\\ \cite{fernandez2023stable,zhao2023recipe,liu2023watermarking,wen2023tree,yang2024gaussian,xiong2023flexible}\end{tabular}                                              &                                                                                &                                                                                       &                                                                                                                                 &                                                                                                                                                            \\ \cline{2-2} \cline{5-6} 
\multicolumn{1}{|c|}{}                                                                                                     & \begin{tabular}[c]{@{}l@{}}Watermarking LLMs \\ \cite{kirchenbauer2023watermark,liu2024adaptive,hu2023unbiased,zhang2023remark,munyer2024deeptextmark,huo2024token}\end{tabular}                                     &                                                                                &                                                                                       & \begin{tabular}[c]{@{}l@{}}$\bullet$ Successful differentiate AI-created\\ content from human-written text\end{tabular}         & \begin{tabular}[c]{@{}l@{}}$\bullet$ Fail to strike a balance between\\ robustness and text quality\end{tabular}                                       \\ \hline
\end{tabular}}
\end{table*}

\subsubsection{Model component-based watermarking} When white box validation is allowed, the model owner has the highest privileges to access the suspect model's structure and parameters, and can even observe the model updating or training process on the test samples. Various model elements, such as model weights, dynamic parameters, and model structures, can be used to embed watermarks.

\par $\bullet$ \textbf{Model weights.} A representative model watermarking embeds watermarks into static weight space by regularizing the objective loss function or reordering weight importance.

\par Early watermarking schemes often chose to incorporate additional terms into the original loss function. Uchida \textit{et al.} \cite{uchida2017embedding} is the first that introduced the watermarking IPP scheme in a white box setting by incorporating the watermark regularizer into the original task loss. Nevertheless, this watermarking drastically modifies the weight distribution and can be easily detected and evaded. 
On this basis, many schemes \cite{wang2021riga,lv2023robustness} have been proposed to enhance covertness and robustness. The Robust whIte box generative adversarial network (GAN) watermarking (RIGA) was introduced by Wang \textit{et al.} \cite{wang2021riga}. It utilizes the concept of adversarial training to guarantee the consistency of weight distributions between the watermarked and original models. 
Lv \textit{et al.} \cite{lv2023robustness} divided a pretrained autoencoder’s weight space into two pieces: embedded piece (i.e., encoder) as the watermark, and secret piece (i.e., decoder) as the local secret. The owner injects the encoder’s weights into the DL model weight space. Two pieces are later combined for IP infringement detection.

\par Another representative watermarking embeds watermarks into either the most or the least significant model weights. Liu \textit{et al.} \cite{liu2021watermarking} proposed an IPP method named greedy residuals, which greedily select fewer yet more salient model weights for embedding, where the residual information is the sum of extracted weight values. The ownership identifier is covertly embedded along with the residual information into the target model. Similarly, Namba \textit{et al.} \cite{namba2019robust} identified model weights in DL models that significantly contribute to predictions and exponentially increase their weight values.

\par $\bullet$ \textbf{Dynamic parameters.} Dynamic components corresponding to inputs, such as activation maps, gradients, and the hidden states of recurrent neural networks (RNNs), can serve as watermark carriers. 
In contrast to model weights that are static, the dynamic components, which rely on both data and the model itself, offer greater flexibility, covertness, and robustness. 
\par Deepsigns \cite{rouhani2019deepsigns} functions by embedding an arbitrary $N$-bit binary string into the probability density function of activation maps across multiple layers. To achieve effective and harmless model watermarking, MOVE \cite{li2022move} hides external features in the gradients of trigger-carrying samples. Gradients of the original and watermarked models are obtained on clean and trigger datasets respectively, and a binary classifier is trained to distinguish the source of gradients. Lim \textit{et al.} \cite{lim2022protect} extended the watermarking to RNNs and embedded it into the hidden states. Here, the protected model is trained using a key matrix $K$, and after each time step, the original hidden state $\bm{\mathit{h}}$ is transformed into $\bm{\mathit{\hat h}}$, which is formulated as follows:
\begin{equation}
\bm{\mathit{\hat h}} = K \otimes \bm{\mathit{h}}, K = BE \odot BC,
\end{equation}
where $BE$ is the binary representation of the owner’s key string; $BC \in {\left\{ { - 1,1} \right\}^{\left| K \right|}}$ is a random sign vector.

\par $\bullet$ \textbf{Model structures.} The model structure can serve as both a carrier for embedding watermarks and as the watermark itself.

\par Inspired by the neural architecture search (NAS) algorithm, Lou \textit{et al.} \cite{xiaoxuan2021meets} proposed to guide NAS in generating sufficiently unique architectures based on owner-specific keys. The architecture itself serves as a watermark for ownership claim while preserving the high usability of the model itself. 
Nonetheless, in practice, it is more common to watermark existing trained models rather than adding watermarks during the process of searching model architectures, as the latter typically requires substantial computational resources. 
Zhao \textit{et al.} \cite{zhao2021structural} segmented the watermark into multiple bit segments and utilized each bit segment to determine pruning rates per convolutional layer. During ownership verification, the watermark can be reliably reconstructed by identifying the channel pruning rates of the suspected models. It is worth noting that this method is only possible for multilayer perceptron (MLP) and convolutional neural network (CNNs). Additionally, the lottery ticket hypothesis (LTH) emerges as a promising framework that explores pruned models without sacrificing model performance. The main idea behind this work \cite{chen2021you} is to retrieve a sparse network while remaining comparable or even better model performance. Any attempt at fine-tuning and further pruning attacks would seriously hinder the model availability because the pruned models in work \cite{chen2021you} are in their most basic form and cannot be further pruned.

\subsubsection{Query-based watermarking} When query-only black box verification is allowed, the model owner can only query some trigger-carrying samples to the suspect model and then observe whether the suspect model’s predictions exhibit a preset output pattern.

\par The workflow of such watermarking schemes can be divided into three stages: (i) generating or selecting trigger-carrying sample pairs $\left( {{x_t},{y_t}} \right)$ to construct a dataset ${\mathcal{D}_T}$; (ii) fine-tuning the model $\Phi$ with the trigger dataset or jointly retraining it with the original dataset $\mathcal{D}$ to embed the watermark. Thereby the watermarked model ${\Phi _{\rm wm}}$  will exhibit preset output patterns; (iii) querying suspect models using the trigger-carrying sample set to obtain predictions and verify the presence of the watermark. According to the means of creating trigger-carrying samples, query-based watermarking schemes are divided into the following four categories:

\par $\bullet$ \textbf{Out-of-distribution (OOD) samples.} Adi \textit{et al.} \cite{adi2018turning} randomly selected OOD samples as trigger-carrying samples. Another factor that can be used to choose trigger-carrying samples is the OOD degree, i.e., the deviation degree from the original data distribution. For instance, Deepsigns \cite{rouhani2019deepsigns} proposed to select samples from sparsely explored regions as trigger-carrying samples based on the probability density function of the model’s activation heatmaps. In~\cite{jia2021entangled}, Jia \textit{et al.} presented an entangled watermark embedding (EWE) method to tweak representations extracted from training samples and trigger-carrying samples, in which the original and trigger data originate from different task distributions.

\par Although these methods can effectively enhance watermarking robustness to some extent, their watermarks are hard to survive in certain common scenarios, such as transfer learning or domain adaptation. A subnetwork-lossless robust DL model watermarking framework named SRDW was suggested in work \cite{jia2022subnetwork}, which intends to enhance IPP during deep transfer learning processes. Using the lottery ticket hypothesis as a guide, SRDW finds the best core subnet for embedding watermarks and employs OOD-guided data augmentation to strengthen the watermarks’ robustness.

\par $\bullet$ \textbf{Samples near to decision boundary.} The prediction of such samples should be deterministic for a given trained model, but it can vary greatly upon different models if the sample is situated close to the DB. 

\par Yang \textit{et al.} \cite{yang2021robust} utilized Shannon entropy (SE) to gauge the inherent uncertainty or confidence in model predictions, further selecting high-entropy samples as keys. It does not rely on end-to-end retraining or fine-tuning key samples with the desired labels. AIME \cite{mehta2022aime} condenses the model’s misclassifications into a confusion matrix and subsequently selects trigger-carrying samples positioned at the DB based on it.

\par $\bullet$ \textbf{Natural samples attached with preset trigger patterns.} Trigger patterns can consist of specific content, such as user signature information, white squares, different image styles, etc. Protected models will operate normally on regular data but exhibit specific behavioral patterns on modified samples.
\par Guo \textit{et al.} \cite{guo2018watermarking} superimposed message masks containing $n$-bit signatures into the original images which are undetectable. Similarly, Li \textit{et al.} \cite{li2019prove} fed the original samples and identity logotypes into the designed encoder framework to generate covert trigger-carrying samples that conceal the owner’s identity. Certified watermarks are proposed to enhance the robustness and irreversibility. Bansal \textit{et al.} \cite{bansal2022certified} introduced random smoothing techniques to ensure the irreversibility of watermarks against ${l_2}$ adversary. Nevertheless, in high-dimensional spaces, this work falls short of offering meaningful robustness proof against ${l_p}(p > 0)$ attacks. Based on partial differential equation theory, Ren \textit{et al.} \cite{ren2023dimension} proposed a mollifier smoothing method to ensure that the certified watermark remains unaffected by high-norm watermark removal attacks ($1 \le p \le \infty $). Further, internal features, such as variations in model predictions for data within the same class, are often unreliable; comparatively, external features are more dependable. To this end, MOVE \cite{li2022move} modified the trigger-carrying samples’ image style and embedded them into the victim model, without changing the label of watermarked samples. Szyller \textit{et al.} \cite{szyller2021dawn} proposed DAWN, which introduces a backdoor mechanism to randomly modify the hard labels of trigger-carrying samples.

\par $\bullet$ \textbf{Synonym replacement.} Synonym replacement can preserve the semantics without requiring significant modifications. Defenders can take advantage of this by replacing some words with the least used or specially selected synonyms, thus stamping the API output with invisible and transferable marks.

\par ToSyn \cite{li2023protecting} is the first watermark designed to protect LLM-based code generation (LLCG) APIs from remote imitation attacks. It is based on the observation that programming languages contain many synonyms at the token level, meaning that replacing tokens with their synonyms does not alter (or only slightly alter) the functionality of the language model’s output. Consequently, IP watermarking is defined as the secret adjustment distribution between synonyms at the token level in the output from LLCG.

\par However, these above methods can be nullified by obvious countermeasures such as ``synonym randomization”. To address this issue, He \textit{et al.} \cite{he2022cater} proposed a CATER framework to inject the watermarks in conditional word distribution, while maintaining the original word distribution. Specifically, an optimization method was designed to determine watermark rules that can minimize overall word distribution distortion while maximizing the variation in conditional word selection. 
Zhao \textit{et al.} \cite{zhao2023protecting} proposed GINSEW, which injects secret signals into the probability vector of the decoding steps for each target token. Setting their sights on pre-trained LLMs, Li \textit{et al.} \cite{li2023plmmark} proposed PLMmark, the first safe and dependable black box watermarking platform. There are two important phases in it: (1) Using the original vocabulary of the model, creating a tight connection between the digital signature and trigger phrases. When this is combined with public key encryption, a watermark with the owner's identity is produced; (2) A supervised contrastive loss is added to guarantee anti-transfer learning, robustness, and generality, resulting in the trigger set's output representations deviating from clean sample representations.

\subsubsection{Generative content-based watermarking} In the most stringent case of black box verification, the defender can only obtain the model-generated content, such as images generated by GANs or DMs. Thus, Watermarks can also be embedded in the outputs according to predefined rules. The application scenario of this kind of solution is generative models. The schemes can be divided into the following categories according to the type of the protected model.

\par $\bullet$ \textbf{Watermarking auto-encoders.}
SSLGuard \cite{cong2022sslguard} draws inspiration from a mathematical proposition: two random vectors in high-dimensional space are nearly orthogonal. Conversely, a notable average cosine similarity, significantly greater than 0 or approaching 1, indicates a strong correlation with these vectors. Defenders can refine a clean encoder to convert trigger-carrying samples ${D_T}$ into embeddings, and then train a decoder to further convert the embeddings into decoding vectors with high cosine similarity to secret vector $sk$.

\par The work \cite{cong2022sslguard} focuses on scenarios where the encoder is deployed separately without downstream classifiers. However, if attackers steal the encoder and customize it to downstream tasks via transfer learning, ownership cannot be verified because the stolen encoder and its embeddings are typically inaccessible. To overcome the challenges, SSL-WM \cite{lv2022ssl} maps watermarked inputs into the encoder’s invariant representation space, which makes any downstream classifier produce the expected behavior. Subsequently, the owner employs an outlier detection algorithm to monitor the downstream classifier's outputs.

\par $\bullet$ \textbf{Watermarking GANs.} 
The simplest watermarking mechanism for GANs involves adding a uniform visible watermark to the output images, which may sacrifice visual quality and usability. Inspired by the above observation, researchers \cite{zhang2021deep,wu2020watermarking, abdelnabi2021adversarial, fei2024wide,lin2024cyclegan} have devised multiple loss terms or multi-objective functions combined with adversarial training to embed invisible model watermarks into the generated-content of GANs. For instance, the work \cite{wu2020watermarking} introduced embedding loss and perceptual loss to ensure visual consistency between the watermarked image and the original image building upon the original task objectives. Additionally, the work \cite{abdelnabi2021adversarial} further incorporated semantics preservation and sentence correctness terms to enforce outputs that are semantically similar to the input sentences and maintain correct grammar and structure. Both works trained an additional watermark-extraction network for subsequent ownership verification.

\par $\bullet$ \textbf{Watermarking DMs.} Recent research endeavors have amalgamated the watermark embedding process with the image generation process of the DMs. The authors in work \cite{fernandez2023stable} fine-tuned the DM decoder using a pre-trained watermark extractor, making it easier to extract watermarks from images. 
Zhao \textit{et al.} \cite{zhao2023recipe} and Liu \textit{et al.} \cite{liu2023watermarking} suggested fine-tuning the DM to implant a backdoor as a watermark. Furthermore, Wen \textit{et al.} \cite{wen2023tree} adapted the frequency domain of latent representations to match specific patterns. However, it directly disrupts the Gaussian distribution of noise, limiting the randomness of sampling and resulting in affecting model performance. Yang \textit{et al.} \cite{yang2024gaussian} observed that the latent representations obtained from non-watermark DMs follow a standard Gaussian distribution. Based on this, Gaussian Shading was developed, which mapped the watermark to latent representations identical to those derived from the non-watermarked distribution. Since the watermark is intricately linked with image semantics, it exhibits resilience to lossy processing and erasure attempts.

\par The methods mentioned above can only embed a fixed message; that is, the message to be embedded cannot be changed unless the model is retrained. The work \cite{xiong2023flexible} proposed an end-to-end watermarking method based on the encoder-decoder and message-matrix. The watermark can be embedded into generated images by fusing the message-matrix and intermediate outputs in the forward propagation. Thus, the watermark message can be flexibly changed by utilizing the message-encoder to generate message-matrix, without training the DM again.

\par $\bullet$ \textbf{Watermarking LLMs.} Note that compared to other data modalities (such as images and audio), text data exhibits a pronounced sparsity. Moreover, there are logical relationships between text tokens; even subtle alterations may confuse or damage the semiotic fidelity.
\par To avoid compromising the semantic coherence of the texts, Zhang \textit{et al.} \cite{zhang2023remark} designed a reparameterization module to change the dense message encoding distribution into a sparse distribution of text tokens with watermarks. A more flexible approach, adaptive text watermarking \cite{liu2024adaptive}, was proposed by Liu \textit{et al.} Specially, the watermark is adaptively added to the high entropy token distribution measured by the auxiliary model and keeps the low entropy token distribution unchanged. The above works validate the trade-off between watermark strength and output quality only from experimental simulations and lack a reliable theoretical analysis. In order to fill this gap, Hu \textit{et al.} \cite{hu2023unbiased} proved that watermarking does not affect the output probability distribution through proper implementation.

\par The above methods are infeasible in practice, as the watermark detector may not have access to the original text. In view of this, Munyer \textit{et al.} \cite{munyer2024deeptextmark} introduced DeepTextMark, a deep learning-driven text watermarking method for text source identification. Watermark insertion is performed by utilizing Word2Vec and sentence encoding, as well as a transformer-based watermark detection classifier. It does not require direct access or changes to the underlying text generation mechanism. Most work faces challenges in simultaneously improving detectability and semantic consistency: improving one often compromises the other. To this end, Huo \textit{et al.} \cite{huo2024token} introduced multi-objective optimization (MOO) that utilizes a lightweight network to generate token-specific watermark logits and segmentation ratios. By using MOO to optimize the detection and semantic objective functions, the presence of watermarks does not affect the model performance in downstream tasks.

\subsection{Model Fingerprinting}
\label{3.3}
According to the objective of model similarity comparison (i.e. model property or model behavior), fingerprints can be divided into two categories. The comparison of these two categories is shown in Table \ref{table:model FP comparison}. 

\begin{table*}[]
\centering
\caption{Comparison of model fingerprinting approaches.}
\label{table:model FP comparison}
\scalebox{0.7}{
\begin{tabular}{|ll|l|c|l|c|l|c|}
\hline
\multicolumn{2}{|c|}{\textbf{Category}}                                                                         & \multicolumn{1}{c|}{\textbf{Papers}}   & \textbf{Year} & \multicolumn{1}{c|}{\textbf{Test Cases}}                                                   & \begin{tabular}[c]{@{}c@{}}\textbf{Reference} \\ \textbf{Models}\end{tabular} & \multicolumn{1}{c|}{\textbf{Test Metrics}}                                                       & \begin{tabular}[c]{@{}c@{}}\textbf{Target}\\ \textbf{Networks}\end{tabular} \\ \hline
\multicolumn{1}{|c|}{\multirow{3}{*}{Static Property}}   & \multirow{2}{*}{Model Parameter}   & \cite{chen2023perceptual}     & 2023 & -                                                                                 & \Checkmark                                                  & \begin{tabular}[c]{@{}l@{}}Model Hash on NTS \\of Critical Weights\end{tabular}        & CNNs                                                      \\ \cline{3-8} 
\multicolumn{1}{|l|}{}                                            &                                    & \cite{xiong2022neural}        & 2022 & -                                                                                 & \Checkmark                                                  & \begin{tabular}[c]{@{}l@{}}Model Hash from \\Hash Generators\end{tabular}              & CNNs                                                      \\ \cline{2-8} 
\multicolumn{1}{|l|}{}                                            & Training Paths                     & \cite{jia2021proof}           & 2021 & Training Samples                                                                  & \XSolidBrush                                                & \begin{tabular}[c]{@{}l@{}}Model Weights\\ Similarity\end{tabular}                     & CNNs                                                      \\ \hline
\multicolumn{1}{|c|}{\multirow{16}{*}{Dynamic Behavior}} & \multirow{7}{*}{Misclassification} & \cite{guan2022you}            & 2022 & \begin{tabular}[c]{@{}l@{}}Wrongly-Classified\\ Samples\end{tabular}              & \XSolidBrush                                                & \begin{tabular}[c]{@{}l@{}}Correlation Matrices\\ of Model Predictions\end{tabular}     & CNNs                                                      \\ \cline{3-8} 
\multicolumn{1}{|l|}{}                                            &                                    & \cite{lukas2019deep}          & 2019 & Conferrable Samples                                                               & \Checkmark                                                  & \begin{tabular}[c]{@{}l@{}}Fingerprint Evaluator\\ on Model Outputs\end{tabular}        & CNNs                                                      \\ \cline{3-8} 
\multicolumn{1}{|l|}{}                                            &                                    & \cite{yang2022metafinger}     & 2022 & \begin{tabular}[c]{@{}l@{}}Triplet Loss Optimized\\Samples\end{tabular}         & \Checkmark                                                  & \begin{tabular}[c]{@{}l@{}}Matching Rate on \\ Hard Predictions\end{tabular}            & CNNs                                                      \\ \cline{3-8} 
\multicolumn{1}{|l|}{}                                            &                                    & \cite{pan2022metav}           & 2022 & \begin{tabular}[c]{@{}l@{}}Adaptive Fingerprints\end{tabular}                  & \Checkmark                                                  & \begin{tabular}[c]{@{}l@{}}Fingerprint Evaluator\\ on Model Outputs\end{tabular}        & CNNs                                                      \\ \cline{3-8} 
\multicolumn{1}{|l|}{}                                            &                                    & \cite{li2020learning}         & 2020 & Adversarial Examples                                                              & \Checkmark                                                  & \begin{tabular}[c]{@{}l@{}}Matching Rate on \\ Hard Predictions\end{tabular}            & CNNs                                                      \\ \cline{3-8} 
\multicolumn{1}{|l|}{}                                            &                                    & \cite{yu2021artificial}       & 2021 & \begin{tabular}[c]{@{}l@{}}Samples Optimized by\\ Linear Programming\end{tabular} & \XSolidBrush                                                & First ReLU Activation                                                                   & CNNs                                                      \\ \cline{3-8} 
\multicolumn{1}{|l|}{}                                            &                                    & \cite{yu2021artificial}       & 2021 & \begin{tabular}[c]{@{}l@{}}Samples Optimized by\\ Linear Programming\end{tabular} & \XSolidBrush                                                & First ReLU Activation                                                                   & CNNs                                                      \\ \cline{2-8} 
\multicolumn{1}{|l|}{}                                            & \multirow{5}{*}{Low Confidence}    & \cite{wang2021fingerprinting} & 2021 & Adversarial Samples                                                               & \Checkmark                                                  & \begin{tabular}[c]{@{}l@{}}Matching Rate on \\ Hard Predictions\end{tabular}            & CNNs                                                      \\ \cline{3-8} 
\multicolumn{1}{|l|}{}                                            &                                    & \cite{cao2021ipguard}         & 2021 & Near-Boundary Samples                                                             & \XSolidBrush                                                & \begin{tabular}[c]{@{}l@{}}Matching Rate on \\ Hard Predictions\end{tabular}            & CNNs                                                      \\ \cline{3-8} 
\multicolumn{1}{|l|}{}                                            &                                    & \cite{chen2022copy}           & 2022 & Adversarial Samples                                                               & \XSolidBrush                                                & \begin{tabular}[c]{@{}l@{}}Three-level Metrics: \\ Property, Neuron, Layer\end{tabular} & CNNs                                                      \\ \cline{3-8} 
\multicolumn{1}{|l|}{}                                            &                                    & \cite{wang2023publiccheck}    & 2023 & Encysted Samples                                                                  & \XSolidBrush                                                & \begin{tabular}[c]{@{}l@{}}Matching Rate on \\ Hard Predictions\end{tabular}            & CNNs                                                      \\ \cline{3-8} 
\multicolumn{1}{|l|}{}                                            &                                    & \cite{peng2022fingerprinting} & 2022 & \begin{tabular}[c]{@{}l@{}}Universal Adversarial\\ Perturbations\end{tabular}     & \Checkmark                                                  & \begin{tabular}[c]{@{}l@{}}Fingerprint Encoder\\ on Soft Predictions\end{tabular}       & CNNs                                                      \\ \cline{2-8} 
\multicolumn{1}{|l|}{}                                            & \multirow{4}{*}{Preset Response}   & \cite{xu2024instructional}    & 2024 & Backdoor Instruction                                                              & \Checkmark                                                  & \begin{tabular}[c]{@{}l@{}}Matching Rate on \\ Model Outputs\end{tabular}               & LLMs                                                      \\ \cline{3-8} 
\multicolumn{1}{|l|}{}                                            &                                    & \cite{iourovitski2024hide}    & 2024 & Adversarial Prompts                                                               & \XSolidBrush                                                & \begin{tabular}[c]{@{}l@{}}Fingerprint Evaluator\\ on Model Outputs\end{tabular}        & LLMs                                                      \\ \cline{3-8} 
\multicolumn{1}{|l|}{}                                            &                                    & \cite{russinovich2024hey}     & 2024 & Adversarial Prompts                                                               & \XSolidBrush                                                & \begin{tabular}[c]{@{}l@{}}Matching Rate on \\Model Outputs\end{tabular}               & LLMs                                                      \\ \cline{3-8} 
\multicolumn{1}{|l|}{}                                            &                                    & \cite{jin2024proflingo}       & 2024 & Preset Prompt Templates                                                           & \Checkmark                                                  & \begin{tabular}[c]{@{}l@{}}Matching Rate on \\Model Outputs\end{tabular}               & LLMs                                                      \\ \hline
\end{tabular}}
\end{table*}

\subsubsection{Static property-based fingerprinting}
The defender can extract the static properties of the model (such as model parameters, training paths) as fingerprints, etc.

\par $\bullet$ \textbf{Model parameter.} The defender can convert the model parameters to hash codes and then verify model ownership by comparing the similarity of hash codes. Chen \textit{et al.} \cite{chen2023perceptual} selected the largest-absolute-values weights and employed model compression techniques to compute normal test statistic for each weight segment to get fixed-length hash codes as fingerprints of the model. 
In work \cite{xiong2022neural}, the model hash sequence consists of two parts: one is the model piracy detection hash based on dynamic outputs of convolution layers; the other is the model tampering localization hash, which can assist model owners in accurately detecting tampered locations.

\par $\bullet$ \textbf{Training paths.} In work \cite{jia2021proof}, a suspect model must furnish a sequence of batch indices and intermediate model weights. The defender selects pairs of model parameter states from the information provided by the suspect to build a validation subset. If the defender fails to use the suspect's model evidence to replicate the trajectory, it indicates illegal derivation from the victim's model. It requires access to a significant portion of the suspect model's raining paths and is fragile to trivial modifications such as model pruning or fine-tuning.

\subsubsection{Dynamic behavior-based fingerprinting} 
The defenders expect that there exists a subclass of targeted, transferable, adversarial fingerprints that can trigger the model-specific knowledge that is shared by source models and their derivations while not shared by any other unrelated models.

\par $\bullet$ \textbf{Misclassification-based.} Test samples with misclassification behavior can be selected either from natural samples or iteratively optimized from an initial data point along the gradient of the objective function. 

\par SAC \cite{guan2022you} leveraged misclassified natural samples as test inputs. Specifically, they introduced mispredicted benign samples and CutMix augmented samples to replace adversarial samples as inputs to the model. Subsequently, a robustness indicator is trained for identifying stolen models. Models that are independently trained and differ from the original model in at least one aspect, such as the training dataset, model structure, training procedure, initial weights, etc., are referred to as negative models. Conversely, models that have been taken from or altered from the protected model are regarded as positive models.
Defenders in CEM \cite{lukas2019deep} employ multiple dropout versions of the source models as positive models. Specifically, CEM \cite{lukas2019deep} searches for a perturbation $\delta$ that produces transferable adversarial examples $x = x + \delta $ by maximizing the output activation difference between the positive and negative models. 
Any transferable examples from the source model should exhibit the same misclassification fidelity in the positive models but different behaviors in the negative models. A limited number of reference models (i.e. positive and negative models) may lead to a decrease in model performance and a high false negative rate. 
To enhance the diversity of the positive and negative models, MetaV \cite{pan2022metav} imports typical model modification methods (e.g., pruning, finetuning, and distillation) to enrich positive models. MetaFinger \cite{yang2022metafinger} and Li \textit{et al.} \cite{li2020learning} designed ghost networks to apply feature-level perturbations to an existing model to potentially create a huge set of positive and negative models.

\par Simply extending fingerprinting techniques from classification tasks to image generative models may pose challenges due to potential vulnerabilities in robustness and covertness. To this end, Yu \textit{et al.} \cite{yu2021artificial} added artificial fingerprints to the training data and enhanced the transferability from the training data to the generated model. However, attackers may find it easier to invalidate its fingerprints by fine-tuning the model or data samples. Further, Li \textit{et al.} \cite{li2021fingerprinting} combined the target GAN with a classifier to construct a composite DL model. The GAN generates fingerprint samples by adding perturbations, which leads to model outputs visually resembling natural ones but mislabeled by the classifier. In the verification phase, the suspicious model outputs are fed into the classifier for ownership verification by observing predictive behavior.

\par Existing fingerprint algorithms frequently necessitate access to training data to generate appropriate test samples as fingerprints. Nonetheless, owing to regulatory constraints, certain model owners lack secondary access rights to the training dataset. Consequently, test cases can be generated from randomly initialized seed samples rather than natural samples. 
For instance, Characteristic Examples \cite{wang2021characteristic} and Intrinsic Examples \cite{wang2021intrinsic} opt to address the misclassification sample generation issue utilizing the projected gradient descent (PGD) algorithm.

\par $\bullet$ \textbf{Low confidence-based.} DL models can be uniquely represented by their DBs. The following design of dynamic fingerprint schemes typically revolves around searching for data points near the DB of the target model.

\par Wang \textit{et al.} \cite{wang2021fingerprinting} exploited the DeepFool algorithm to generate samples that are only a minimal deviation between the maximum probability and the second maximum probability of sample predictions. IPGuard \cite{cao2021ipguard} and DeepJudge \cite{chen2022copy} treated the search for fingerprint samples as an optimization problem, in which the PGD algorithm was used to steer the starting data points toward DBs. In a dual black box scenario, PublicCheck \cite{wang2023publiccheck} generated smoothly transformed and augmented encysted samples that are enclosed around the model’s DB while ensuring that the verification queries are indistinguishable from normal queries. Peng \textit{et al.} \cite{peng2022fingerprinting} adds a universal adversarial perturbation vector on the test cases and adopts $k-$means clustering on the last layer of the victim model to ensure that the datapoints move towards multi-target classes.

\par $\bullet$ \textbf{Preset response-based.} Pre-trained LLMs can be fine-tuned in countless ways to adapt downstream tasks \cite{diwan2021fingerprinting}. Consequently, the copyright protection for LLMs faces a new challenge: ``Is this suspect model fine-tuned from the source LLMs?"

\par Iourovitski \textit{et al.} \cite{iourovitski2024hide} designed a novel ``Hide and Seek” approach in which a detective LLM uses the preset responses to fingerprint the target models while an auditor LLM creates discriminative prompts. Nevertheless, this method demands expensive training and relies on prior knowledge of user downstream tasks or datasets. To this end, works \cite{xu2024instructional} and \cite{russinovich2024hey} identified a collection of multi-dimensional metrics that an effective fingerprint mechanism should have in light of the shortcomings of the previously described techniques. They then painstakingly built a fingerprinting scheme that can strike a delicate balance between these crucial metrics. Xu \textit{et al.} \cite{xu2024instructional} provided a preliminary study on LLM fingerprinting utilizing lightweight instruction tweaking. Detailed, the model publisher implanted a secret private key as an instruction backdoor, causing the LLM to produce preset texts when the key is present. The Chain \& Hash implemented a fingerprint with a cryptographic flavor \cite{russinovich2024hey}. It entails coming up with a list of queries (the fingerprints) and a list of possible responses. To provide an unforgeability property, these components are hashed together using a secure hashing technique to choose the value for each question. Similarly, ProFLingo \cite{jin2024proflingo} generated queries that elicit specific responses from an original model, thereby establishing unique fingerprints.

\section{Proactive Model Intellectual Property Protection}
\label{3.1}
\subsection{Overview}
\begin{figure}[htbp]
\vspace{-0.1in}
	\centering
\includegraphics[width=.9\textwidth]{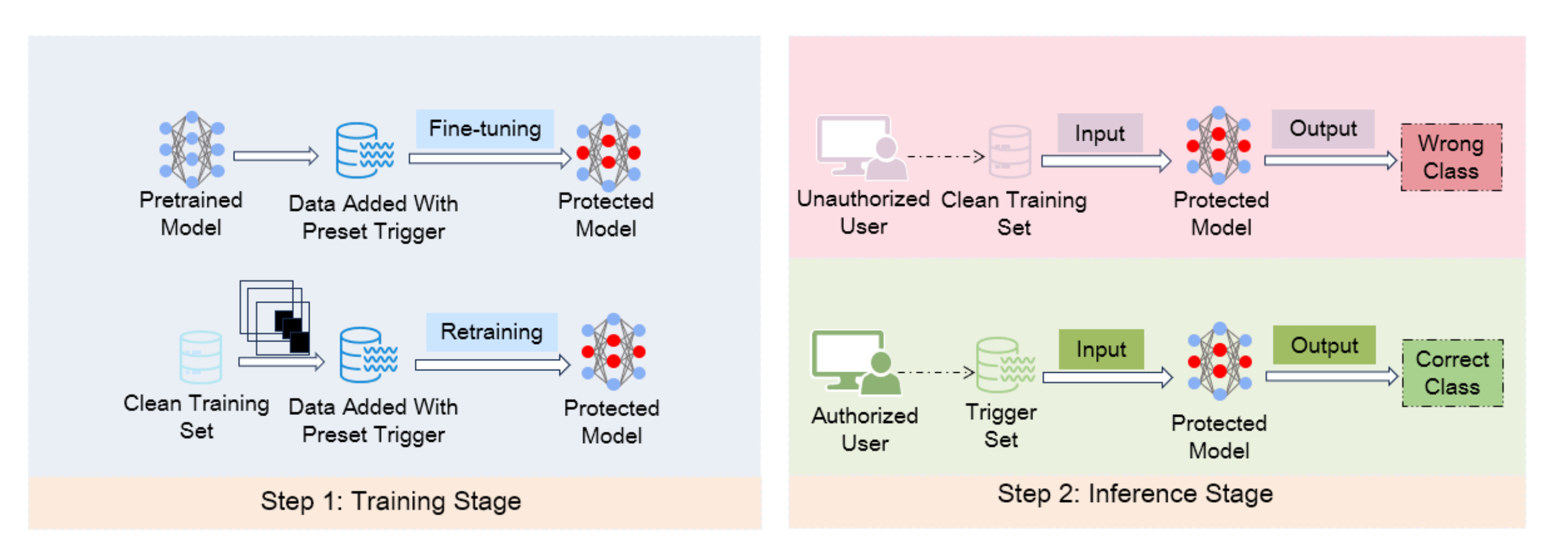}
 \vspace{-0.1in}
	\caption{The overall flow of the proactive IPP method.}
	\label{fig:proactive overall}
 \vspace{-0.1in}
\end{figure}

The reactive schemes outlined above claim ownership after the infringement has already occurred, which is both delayed and insufficient. 
To prevent model theft proactively, an increasing number of researchers are focusing on proactive IPP. The overall flow of the proactive IPP method is presented in Fig. \ref{fig:proactive overall}. According to the management perspective, the proactive IPP has the following types:

\begin{itemize}
\item \textit{Proactive authorization control}: The proactive defenders focus on maintaining the DL model's functionality for authorized users while rendering them dysfunctional for unauthorized users. Moreover, user identity management and authorization control can be further combined in order to track betrayers when model infringement occurs.

\item \textit{Domain authorization control}: Comprehensive IPP necessitates thorough consideration of domain authorization control in order to prevent authorized users from transferring the model to any tasks without restriction, which potentially leads to implicit infringement.
\end{itemize}

\par Overall, proactive authorization control is suitable for directly preventing unauthorized use, while domain authorization control is more concerned with limiting the scope of authorized use. Each of them has unique advantages in their respective protection layers. The summary of these two categories of proactive IPP is shown in Table \ref{table:proactive model IPP summary}.

\begin{table*}[]
\centering
\caption{Summary of proactive model IPP approaches.}
\label{table:proactive model IPP summary}
\scalebox{0.7}{
\begin{tabular}{|c|l|c|c|c|l|l|}
\hline
\textbf{Paper}                                             & \multicolumn{1}{c|}{\textbf{Year}} & \textbf{\begin{tabular}[c]{@{}c@{}}Traitor\\ Traceable\end{tabular}} & \textbf{\begin{tabular}[c]{@{}c@{}}Domain\\ Restriction\end{tabular}} & \textbf{\begin{tabular}[c]{@{}c@{}}Construction\\ Scenarios\end{tabular}} & \multicolumn{1}{c|}{\textbf{\begin{tabular}[c]{@{}c@{}}MLaaS\\ Mode\end{tabular}}} & \multicolumn{1}{c|}{\textbf{Drawbacks}}                                                                                                                                     \\ \hline
Securenet \cite{li2024securenet}                            & \multicolumn{1}{c|}{2024}          & \XSolidBrush                                                         & \XSolidBrush                                                          & Retraining                                                                & \begin{tabular}[c]{@{}l@{}}White \&\\ Black box\end{tabular}                       & \multirow{2}{*}{\begin{tabular}[c]{@{}l@{}}Trigger tokens based on backdoor can lead to the\\ misidentification of clean inputs.\end{tabular}}                    \\ \cline{1-6}
Wu \textit{et al.} \cite{wu2022sample}                     & 2022                               & \XSolidBrush                                                         & \XSolidBrush                                                          & Fine-tuning                                                               & Black box                                                                          &                                                                                                                                                                             \\ \hline
Chen \textit{et al.} \cite{chen2018protect}                & 2018                               & \XSolidBrush                                                         & \XSolidBrush                                                          & Retraining                                                                & Black box                                                                          & \begin{tabular}[c]{@{}l@{}}Adversarial perturbations inevitably compromise the\\ model utility.\end{tabular}                                                                 \\ \hline
AprilPyone \textit{et al.} \cite{maungmaung2021protection} & 2021                               & \XSolidBrush                                                         & \XSolidBrush                                                          & Retraining                                                                & \begin{tabular}[c]{@{}l@{}}White \&\\ Black box\end{tabular}                       & \begin{tabular}[c]{@{}l@{}}Apply the same transformation mechanism is difficult \\ to combat collusion attacks.\end{tabular}                                                \\ \hline
Luo \textit{et al.} \cite{luo2021hierarchical}             & 2021                               & \XSolidBrush                                                         & \XSolidBrush                                                          & Fine-tuning                                                               & Black box                                                                          & Lack the capability against various advanced attacks.                                                                                                                       \\ \hline
M-LOCK \cite{ren2022protecting}                            & 2022                               & \XSolidBrush                                                         & \XSolidBrush                                                          & Retraining                                                                & Black box                                                                          & \begin{tabular}[c]{@{}l@{}}Transforming the decision boundary can easily leading \\to a decrease in classification accuracy.\end{tabular}                     \\ \hline
ChaoWs \cite{lin2020chaotic}                               & 2021                               & \XSolidBrush                                                         & \XSolidBrush                                                          & Fine-tuning                                                               & White box                                                                          & Introduce high calculation overhead.                                                                                                                                        \\ \hline
DeepIPR \cite{fan2021deepipr}                              & 2021                               & \XSolidBrush                                                         & \XSolidBrush                                                          & Fine-tuning                                                               & White box                                                                          & \multirow{2}{*}{Vulnerable to overwriting attacks.}                                                                                                                         \\ \cline{1-6}
EdgePro \cite{chen2024edgepro}                             & 2024                               & \XSolidBrush                                                         & \XSolidBrush                                                          & Retraining                                                                & Black box                                                                          &                                                                                                                                                                             \\ \hline
AdvParams \cite{xue2022advparams}                          & 2022                               & \XSolidBrush                                                         & \XSolidBrush                                                          & Fine-tuning                                                               & White box                                                                          & Fail to resist collusive attacks by malicious users.                                                                                                                        \\ \hline
NNSplitter \cite{zhou2023nnsplitter}                       & 2023                               & \XSolidBrush                                                         & \XSolidBrush                                                          & Fine-tuning                                                               & Black box                                                                          & \begin{tabular}[c]{@{}l@{}}The computation of the confusion matrix introduces \\ additional computational overhead.\end{tabular}                                            \\ \hline
DeepMarks \cite{chen2019deepmarks}                         & 2019                               & \Checkmark                                                           & \XSolidBrush                                                          & Retraining                                                                & White box                                                                          & It brings a large calculation overhead.                                                                                                                                     \\ \hline
DSN \cite{tang2023deep}                                    & 2023                               & \Checkmark                                                           & \XSolidBrush                                                          & Fine-tuning                                                               & White box                                                                          & Lack of robustness against IP infringement attacks.                                                                                                                         \\ \hline
Xue \textit{et al.} \cite{xue2020active}                   & 2020                               & \Checkmark                                                           & \XSolidBrush                                                          & Retraining                                                                & White box                                                                          & \begin{tabular}[c]{@{}l@{}}The boundaries between high and medium confidence\\ are fuzzy.\end{tabular}                                                                      \\ \hline
Wang \textit{et al.} \cite{wang2022buyer}                  & 2022                               & \Checkmark                                                           & \XSolidBrush                                                          & Retraining                                                                & Black box                                                                          & Lack of robustness against IP infringement attacks.                                                                                                                         \\ \hline
ActiveGuard \cite{xue2023activeguard}                      & 2023                               & \Checkmark                                                           & \XSolidBrush                                                          & Retraining                                                                & White box                                                                          & \multirow{2}{*}{\begin{tabular}[c]{@{}l@{}}Triggers are OOD distributed, which can compromise\\ the model performance.\end{tabular}}                                        \\ \cline{1-6}
DeepAuth \cite{lao2022deepauth}                            & 2022                               & \Checkmark                                                           & \XSolidBrush                                                          & Retraining                                                                & White box                                                                          &                                                                                                                                                                             \\ \hline
Wang \textit{et al.} \cite{wang2021non}                    & 2021                               & \XSolidBrush                                                         & \Checkmark                                                            & Retraining                                                                & \begin{tabular}[c]{@{}l@{}}White \&\\ Black box\end{tabular}                       & The convergence region of NTL is not tight enough.                                                                                                                          \\ \hline
Zeng \textit{et al.} \cite{zeng2022unsupervised}           & 2022                               & \XSolidBrush                                                         & \Checkmark                                                            & Retraining                                                                & \begin{tabular}[c]{@{}l@{}}White \&\\ Black box\end{tabular}                       & Require access to source dataset.                                                                                                                                           \\ \hline
CUTI-domain \cite{wang2023model}                           & 2023                               & \XSolidBrush                                                         & \Checkmark                                                            & Retraining                                                                & \begin{tabular}[c]{@{}l@{}}White \&\\ Black box\end{tabular}                       & \begin{tabular}[c]{@{}l@{}}Require access to source data when performing IP \\ protection, rendering it unsuitable for decentralized\\ private data scenarios.\end{tabular} \\ \hline
SDPA \cite{wang2023domain}                                 & 2023                               & \XSolidBrush                                                         & \Checkmark                                                            & Retraining                                                                & \begin{tabular}[c]{@{}l@{}}White \&\\ Black box\end{tabular}                       & \begin{tabular}[c]{@{}l@{}}The uncertain divergence ball boundary may damage\\ the model utility of the source domain to some extent.\end{tabular}                   \\ \hline
MAP \cite{peng2024map}                                     & 2024                               & \XSolidBrush                                                         & \Checkmark                                                            & Fine-tuning                                                               & White box                                                                          & The convergence region is inaccurate.                                                                                                                                         \\ \hline
\end{tabular}}
\end{table*}

\subsection{Proactive Authorization Control without Tracking}
It involves distinguishing between authorized and unauthorized users and providing different functionalities and performance based on the user’s identity. This is achieved through a variety of techniques, mainly relying on techniques such as input ``encryption” \cite{li2024securenet,wu2022sample,maungmaung2021protection,luo2021hierarchical,luo2021hierarchical}, model ``encryption”\cite{lin2020chaotic,fan2021deepipr,chen2024edgepro}, trusted execution environments (TEE)-shielded protection \cite{hou2021model,tramer2018slalom,asvadishirehjini2022ginn,hashemi2021darknight}.

\par \textit{1) Input ``encryption”.} The defender can identify the user’s identity based on whether the input contains ``keys” (i.e. trigger-carrying samples). The DL model produces poor accuracy if a specific ``key” is absent, while it maps only the ``keyed” inputs into correct predictions.

\par Li \textit{et al.} \cite{li2024securenet} introduced an access license framework named SecureNet. Visual and invisible keys for authorized users were designed: the former replaces specific pixel values with a unique pattern, while the latter generates noise images and implicitly overlays them with the original inputs. In contrast to work \cite{li2024securenet}, the sample-specific transformation method \cite{wu2022sample} incurs lower additional costs, which utilizes a U-Net network to convert clean labels into target labels before fine-tuning the trained DL model. Inspired by adversarial samples, Chen \textit{et al.} \cite{chen2018protect} devised a conversion module that introduces adversarial perturbation to the inputs. Consequently, only authorized users can obtain accurate model predictions, as anti-piracy DL models exclusively recognize these modified inputs. Nevertheless, the introduction of noise or target labels can eventually cause data availability to be disrupted, which lowers classification performance.

\par In a similar vein, AprilPyone \textit{et al.} \cite{maungmaung2021protection} implemented block-wise alterations on input images with a secret key, which employs techniques like pixel shuffling. Even in cases where attackers gain white box access to the model, they cannot use it without the corresponding secret key. Model retraining encounters challenges when utilizing the techniques outlined in \cite{chen2018protect,maungmaung2021protection}. The model’s learning complexity is heightened by input ``encryption” operations, leading to sluggish and unstable convergence. A revolutionary model locking (M-LOCK) scheme was proposed by Ren \textit{et al.} \cite{ren2022protecting} to achieve built-in proactive defense. Backdoor triggers are implanted into the data of authorized users as keys, and the labels of unauthorized users are poisoned. The model only maps inputs with triggers to correct predictions and maximizes the mutual information between unauthorized inputs and poisoned labels. M-LOCK is suitable for specific scenarios, particularly tasks with offline requirements. It can seamlessly integrate into neural networks and be deployed in offline settings to implement data-level validation mechanisms.

\par \textit{2) Model ``encryption”.} The operations can target the convolutional kernels, BatchNorm layer, neurons, or model weights. These kinds of methods are almost reversible and have no negative impact on classification accuracy.
\par Lin \textit{et al.} \cite{lin2020chaotic} proposed a framework called ChaoWs based on the chaotic map theory to encrypt the convolutional kernels. To get the right prediction back from the model, users must purchase a key to decode it. DeepIPR \cite{fan2021deepipr} embeds scale factors (passports) into the BatchNorm layer to resist malicious attacks. It intends to adjust the inference performance based on the passport provided by the user, i.e., the performance will deteriorate significantly due to altered and forged passports. 
Chen \textit{et al.} \cite{chen2024edgepro} proposed a practical, and general edge device model protection method at neuron level, denoted as EdgePro. This method selects a subset of neurons at each layer, replacing their activation values with a locked value. The works \cite{lin2020chaotic,fan2021deepipr,chen2024edgepro} are lightweight defense methods that avoid model retraining, incur low computational overhead, and require minimal modification. 
\par Xue \textit{et al.} \cite{xue2022advparams} designed a weight encryption technique based on adversarial perturbations, named AdvParams, where the positions of encrypted weights and perturbation values form a key. Similar to this, Zhou \textit{et al.} \cite{zhou2023nnsplitter} presented NNSplitter, an automatic weight obfuscation method in which the indices of the obfuscated weights and their original values comprise the model secret. Note that, despite the complex matrix computations involved in encrypting and recovering model weights, works \cite{xue2022advparams,zhou2023nnsplitter} only require operating a small fraction of the weights, thereby notably reducing decryption time.

\par \textit{3) TEE-shielded DL model protection.} TEE-shielded DL model partition (TSDP) techniques offload privacy-insensitive portions to the GPU while shielding privacy-sensitive parts within the TEEs. The model parts are typically partitioned from two perspectives: model structure (e.g., layers) or computation content (e.g., confidence scores, loss, and gradients).
\par $\bullet$ \textbf{Model structure-based TSDP.} Many prior works \cite{hanzlik2021mlcapsule} have explored shielding the whole DL model directly within TEEs, but suffer from extremely high latency (up to $50 \times$) due to the limited computing resources on the CPU’s enclave.

\par To avoid the above problems, defenders can partition DL models based on layer depth, such as placing deep layers \cite{mo2020darknetz}, shallow layers, non-linear layers \cite{ng2021goten}, or intermediate layers \cite{xiang2021aegisdnn} in TEEs. For instance, DarkneTZ \cite{mo2020darknetz} considered that deep layers have a higher probability of leaking private data information and thus executed these layers within TEEs. To ensure timely task execution while selecting the right set of layers for protection, AegisDNN \cite{xiang2021aegisdnn} designed a dynamic programming-based algorithm, which employs a layer-wise DL model timing and silent data corruption (SDC) analysis mechanism to determine a layer protection configuration for each task. TEESLICE \cite{zhang2023no} employs a ``partition-then-train” strategy. It first partitions the DL model into a backbone and several private slices. The public pre-trained model serves as the backbone, while private data trains the slices within the TEE. Slices are obtained by the carefully designed dynamic pruning algorithm, starting with large slices for maximum accuracy, and then optimizing their size under a precision loss threshold.

\par $\bullet$ \textbf{Computation content-based TSDP.} Matrix multiplication (MM) is used in DL models for both forward and backward propagation of computations in convolutional and fully connected layers, resulting in large computational resource consumption. The motivation behind such TSDP methods is to securely outsource MM to GPUs \cite{hou2021model,tramer2018slalom}. For instance, the GPU carries out forward and backward propagation on mini-batches and communicates the computed gradients to the TEE. After that, the TEE updates the weight and clips the gradients \cite{asvadishirehjini2022ginn}. Besides the model partitioning, security strategies must be designed to guarantee both data and model integrity. For instance, DarKnight \cite{hashemi2021darknight} employs a customized data encoding strategy based on matrix masking to create input obfuscation within the TEE. The obfuscated data is then offloaded to the GPU for fast linear algebra computation.

\subsection{Proactive Authorization Control with Tracking}
The previously discussed proactive authorization control schemes are vulnerable to attacks initiated by dishonest users, such as collusion attacks. The following works focus on embedding unique user identity keys for traitor tracking. The classification of these three categories is mainly based on the different carriers of user identity information.

\par \textit{1) Model weights.} This category generates a unique binary identity key for each authorized user and embeds it into the model weight space. The first anti-collusion security framework, called DeepMarks, was presented by Chen \textit{et al.} \cite{chen2019deepmarks}. For each user, it creates a unique $n$-bit binary code and then embeds code into the host network weights. 
However, as the weight space containing identity information is redundant rather than essential, adversaries can readily remove it through fine-tuning. Tang \textit{et al.} \cite{tang2023deep} introduced termed deep serial number (DSN) to produce unique serial numbers as unique binary identity keys for each authorized user. It first trains a teacher model and then utilizes knowledge distillation to transfer its knowledge to student models. Each student model has a unique serial number assigned to it during the distillation process. The user can only utilize the proper model if they supply a valid serial number. The designed DSN exhibits sufficient robustness when addressing various attack methods, such as model pruning, and watermark overwriting. This method seems costly because it requires to create a student model for each user.

\par \textit{2) Backdoor samples.} The user’s identity key can uniquely correspond to the configured backdoor trigger pattern. Xue \textit{et al.} \cite{xue2020active} presented a user management framework based on multi-trigger backdoors, which incorporates $N$ sub-backdoors into the DL model. Each authorized user is assigned with $n$ ($n$$<$$N$) moderate-confidence backdoor signals as a unique identity key. To access the well-trained model, the user must present his backdoor instance set containing $n$ backdoor signals. If the model exhibits specific behavior with medium confidence on his dataset, the user is deemed legitimate and granted access. A buyer-traceable DL model protection was developed by Wang \textit{et al.} \cite{wang2022buyer}. To build watermark samples, defenders augment clean samples with dirty labels. Each buyer is associated with a specific trigger. The model must be fine-tuned by the backdoor trigger set before it is accessible. The defense efficacy of the techniques in works \cite{xue2020active,wang2022buyer} suffers from a high false-positive rate in ownership verification.

\par \textit{3) Adversarial examples.} The model's different response behavior to adversarial samples can be a powerful tool for tracking user identity. ActiveGuard \cite{xue2023activeguard} leveraged well-crafted adversarial examples with specific classes and confidence to serve as users’ fingerprints. Unauthorized users will experience a significant reduction in model performance. DeepAuth \cite{lao2022deepauth} treats the model predictions of adversarial samples as the unique and fragile signature for each protected DL model. This design aims to produce adversarial samples near the DB by calibrating the latent space and aligning gradient directions.

\subsection{Domain Authorization Control}
For authorization control, prior solutions focus on granting user permission to utilize the model; nevertheless, authorized users retain the freedom to apply the model to any application domain. 
Consequently, several studies have introduced domain authorization schemes that aim to substantially degrade the well-trained model performance on unauthorized data domains. According to the defender’s concurrent access to both source training data and target unauthorized data, we classify existing application authorization methods into the following three types:

\par \textit{1) Target-specified model IPP.} The defender has access to both the original training data and the target domain data.
\par Wang \textit{et al.} \cite{wang2021non} proposed target-specified non-transferable learning (NTL) method, which builds an estimator with the characteristic kernel from reproduction kernel hilbert spaces. This estimator aims to extract nuisance-dependent representations by enlarging the maximum average discrepancy between the data distributions from the source and target domains. The work \cite{zeng2022unsupervised} proposed a new unsupervised NTL (UNTL) method, extending the aforementioned target-specified approach discussed in work \cite{wang2021non} to scenarios where the target data is unlabeled. Besides, a secret key component is further introduced to restore the classification capability of the protected model on the target domain.

\par \textit{2) Source-only model IPP.} The defender can only access the original training data and is unaware of the target domain where the malicious attacker performed the model transfer.
\par When the unauthorized target domain is unknown, directly feeding the target domain and original domain into the network for model training is infeasible. To solve this, the defender can construct simulated target domains. For example, Wang \textit{et al.} \cite{wang2023model} inserted Gaussian noise into the adaptive instance normalization (AdaIN) method based on GAN \cite{zhang2024rethink} to generate synthesized samples in place of the target domain train set. Inspired by distributional robust statistics, a lightweight method called domain-specific optimization \cite{wang2023domain} achieves this by defining a divergence ball centered on the training distribution, covering all nearby distributions around the training domain. During the model training phase, the defenders' goals in works \cite{wang2023model} and \cite{wang2023domain} are to minimize the loss within the training domain while maximizing that for other domains.

\par \textit{3) Data-free model IPP.} The defender is unable to access the data in both the original and target domains.
\par Regardless of promising results in \cite{wang2021non,zeng2022unsupervised,wang2023model,zhang2024rethink,wang2023domain}, these methods require concurrent access to both source and target data when performing IPP, rendering them unsuitable for decentralized private data scenarios. To this end, Peng \textit{et al.} \cite{peng2024map} introduced mask pruning (MAP), a framework designed for scenarios where only a well-trained source model is available. They hypothesize that well-trained models have parameters specific to the target domain, and pruning these is the key to protecting IP while possessing integrity. MAP iteratively optimizes a binary mask matrix to maximize risk on the target domain while minimizing it on the source domain. Due to the lack of knowledge from the source and target domains, its performance slightly declines but remains within an acceptable range.

\section{Dataset Intellectual Property Protection}
\label{4}

\subsection{Overview}
\par IPP has overwhelmingly focused on protecting well-trained models and verifying the model creator’s identity. Nevertheless, the DL model will infringe on the dataset's ownership if trained on a dataset without authorization. In such cases, model IPP methods fail to establish a clear link between the DL model and the protected training data. How to verify the ownership of the target data with respect to a suspect ML model remains largely open. The protected data can have diverse modalities; they can be text, pictures, non-fungible tokens, radio data, GPS data, etc. The summary of the dataset IPP approaches is in Table \ref{table:dataset IPP summary}.

\begin{table*}[]
\centering
\caption{Summary of dataset IPP approaches.}
\label{table:dataset IPP summary}
\scalebox{0.7}{
\begin{threeparttable}
\begin{tabular}{|c|l|l|c|c|c|}
\hline
\textbf{Category}                                                                                                                                        & \multicolumn{1}{|c|}{\textbf{Advantages}}                                                                                                                           & \multicolumn{1}{|c|}{\textbf{Drawbacks}}                                                                                                                        & \textbf{\begin{tabular}[c]{@{}c@{}}Resource\\ Cost\end{tabular}} & \textbf{\begin{tabular}[c]{@{}c@{}}Model\\ Fidelity\end{tabular}} & \textbf{Robustness}          \\ \hline
\begin{tabular}[c]{@{}c@{}}Backdoor\\ Watermarking\\ \cite{li2023black,li2020open,ren2024you,li2022untargeted,tang2023did,liu2023copyright}\end{tabular} & \begin{tabular}[c]{@{}l@{}}The use of backdoor attacks to embed \\external patterns is straightforward and \\easy to implement.\end{tabular} & \begin{tabular}[c]{@{}l@{}}The backdoor-based dataset watermarks\\ can never achieve truly harmless\\ verification.\end{tabular}          & \CIRCLE\LEFTcircle\Circle\Circle                                     & \CIRCLE\LEFTcircle\Circle\Circle                                     & \CIRCLE\CIRCLE\Circle\Circle \\ \hline
\begin{tabular}[c]{@{}c@{}}Domain\\ Watermarking\\ \cite{guo2024domain}\end{tabular}                                                                     & \begin{tabular}[c]{@{}l@{}}The functionality of the model is maintained\\ and the covertness of the watermark\\ is guaranteed.\end{tabular}       & \begin{tabular}[c]{@{}l@{}}The production of hardly-generalized\\ samples is computationally complex \\ and expensive.\end{tabular}       & \CIRCLE\CIRCLE\CIRCLE\Circle                                     & \CIRCLE\CIRCLE\CIRCLE\LEFTcircle                                     & \CIRCLE\CIRCLE\CIRCLE\LEFTcircle \\ \hline
\begin{tabular}[c]{@{}c@{}}Function\\ Watermarking\\ \cite{liu2023watermarking,li2023functionmarker}\end{tabular}                                          & \begin{tabular}[c]{@{}l@{}}Knowledge injection does not compromise\\ the model functionality.\end{tabular}                                          & \begin{tabular}[c]{@{}l@{}}It is difficult to resist IP removal attacks.\end{tabular}                                                   & \CIRCLE\Circle\Circle\Circle                                     & \CIRCLE\CIRCLE\CIRCLE\CIRCLE                                     & \CIRCLE\LEFTcircle\Circle\Circle \\ \hline
\begin{tabular}[c]{@{}c@{}}Prediction \\Margin-based\\Fingerprinting\\   \cite{maini2021dataset,20220407800,dziedzic2022dataset}\end{tabular}                                    & \begin{tabular}[c]{@{}l@{}}Selecting the inherent knowledge as the\\ fingerprint perfectly preserves the model \\functionality.\end{tabular}       & \begin{tabular}[c]{@{}l@{}}Suffer from a high false-negative rate.\end{tabular}                                                     & \CIRCLE\CIRCLE\Circle\Circle                                     & \CIRCLE\CIRCLE\CIRCLE\LEFTcircle                                    & \CIRCLE\CIRCLE\Circle\Circle \\ \hline
\begin{tabular}[c]{@{}c@{}}Prediction \\Behavior-based\\Fingerprinting\\ \cite{liu2022your}\end{tabular}                                                   & \begin{tabular}[c]{@{}l@{}}Establish a clear association between the ML \\model and training data.\end{tabular}                               & \begin{tabular}[c]{@{}l@{}}The samples that have a similar influence \\on the prediction behavior of different ML\\ model are hard to find.\end{tabular} & \CIRCLE\CIRCLE\CIRCLE\Circle                                     & \CIRCLE\CIRCLE\CIRCLE\Circle                                     & \CIRCLE\CIRCLE\LEFTcircle\Circle \\ \hline

\begin{tabular}[c]{@{}c@{}}Dataset \\ Authorization\\ Control\\ \cite{xue2023dataset}\end{tabular}                                                        & \begin{tabular}[c]{@{}l@{}}Prevent the occurrence of dataset infringement.\end{tabular}                                                     & It can't track down traitors.                                                                                                             & \CIRCLE\CIRCLE\CIRCLE\LEFTcircle                                    & \CIRCLE\CIRCLE\CIRCLE\LEFTcircle                                     & \CIRCLE\CIRCLE\LEFTcircle\Circle \\ \hline
\end{tabular}
\begin{tablenotes}
\centering
\footnotesize     
\item The more $\CIRCLE$ indicates that the approach proposed in the work is more capable in this aspect.
\end{tablenotes}
\end{threeparttable}}
\end{table*}

\subsection{Dataset Watermarking}
\par According to the crafting means of watermark samples, can be subdivided into the following three categories:

\par \textit{1) Backdoor watermarking.} DVBW \cite{li2023black} and \cite{li2020open} exploit poison-only backdoor attacks to embed external patterns for dataset watermarking so that protected models would have defender-preset prediction behaviors. However, this method tends to introduce new security risks, such as watermark forgery. Beyond classification tasks, such approaches can extend to visual prompt learning, which focuses on learning input perturbations—a visual prompt—added to downstream task data for prediction. However, a significant drawback is that prompts can be easily replicated and redistributed.
To mitigate this, Ren \textit{et al.} \cite{ren2024you} proposed a method named WVPrompt, which leverages pure poison backdoor attack techniques to embed watermarks into the prompts. Further, Li \textit{et al.} \cite{li2022untargeted} investigated how to perform innocuous and imperceptible dataset ownership verification by employing untargeted backdoor watermarks. The work \cite{tang2023did} introduced a clean-label backdoor watermarking framework, which substituted imperceptible adversarial perturbations for mislabeled samples in order to balance the watermark robustness and data distortion. Large training datasets can be refined into smaller ones through dataset distillation, all the while guaranteeing that models trained on both datasets have identical test accuracy. As a result, dataset distillation has considerable economic value and saves customers a lot of money and effort during training. To this end, Li \textit{et al.} \cite{liu2023copyright} were devoted to protecting IP during the dataset distillation process. More specifically, they created backdoor samples with particular markers using a key, thereby embedding the model’s watermark knowledge into the synthesized tiny dataset. 

\par \textit{2) Domain watermarking.} Inspired by the generalization property of DL models, Guo \textit{et al.} \cite{guo2024domain} proposed the first non-backdoor-based dataset IPP method named domain watermark. 
The intuition is to find a hardly generalized domain for the original dataset as its triggers and make the watermarked models correctly classify these triggers that the benign model will misclassify. Specifically, to guarantee watermark covertness, the hardly-generalized domain was specifically constructed as a bi-level optimization, and a collection of visually indistinguishable clean-label changed data with effects similar to domain-watermarked samples was generated. In contrast to backdoor watermarking, domain watermarking tends to be a harmless and covert IPP method.

\par \textit{3) Function watermarking.} Inspired by the backdoor and membership inference attacks in NLP models, Liu \textit{et al.} \cite{liu2023watermarking} proposed Textmarker to protect the classification language dataset. 
In this method, specific triggers (e.g., ``Ops” and ``Less is more”) are added to the text for ownership verification. 
However, backdoor-based methods might affect the semantic information of the watermarked text and perform poorly on non-classification datasets, such as code generation datasets. To address these issues, Li \textit{et al.} \cite{li2023functionmarker} proposed FunctionMarker, which assumed that the attacker could access the target dataset. In the watermarking generation stage, the defender encodes watermark texts into ASCII codes, which are subsequently embedded into the coefficients of customizable functions.
After obtaining the function watermarking, the data owner needs to design prompt-response pairs to enable LLM to learn the watermark function in the process of supervised fine-tuning. In the watermarking extraction stage, the data owner can query the suspect LLMs to extract the pre-defined watermark based on the designed prompts.

\subsection{Dataset Fingerprinting}
\par Dataset fingerprints, as opposed to model fingerprints, seek to locate intermediates that show the inherent connection between a particular model and the training dataset that goes along with it. According to the intermediary generation technologies, dataset fingerprinting can be divided into the following two categories:

\par \textit{1) Prediction margin-based.} Irrespective of the attack method used, the adversary’s goal is always to acquire the knowledge contained in the training set or its derivatives. This enables the dataset owner to introduce prediction margin-based defense.
\par The key intuition in work \cite{maini2021dataset} is that the prediction margin distribution (i.e., distance from the DB) of any pirated model will resemble that of the victim model. Therefore, the victim can assert ownership by comparing the similarity of prediction boundaries between the two models on test samples.
However, Szyller \textit{et al.} \cite{20220407800} conducted extensive robustness evaluation experiments on dataset inference (DI) algorithm in work \cite{maini2021dataset}. The results indicate that DI suffers from a high false-negative rate, as it tends to erroneously identify independently trained models from non-overlapping datasets within the same distribution. In addition, the DI algorithm \cite{maini2021dataset} cannot be applied to self-supervised models, such as encoders because it relies on calculating the distance between data points and DBs, which do not exist in SSL encoders. 
To fill this gap, Dziedzic \textit{et al.} \cite{dziedzic2022dataset} proposed training a density estimation model to claim ownership. If an encoder is trained on the victim's training data, the log-likelihood of the encoder's output representations is higher compared to test data.

\par \textit{2) Prediction behavior-based.} Liu \textit{et al.} \cite{liu2022your} have observed that a part of the data has a similar influence on the prediction behavior of different ML models. As a result, by comparing the prediction behaviors of the suspect model and the model trained on the target data, the model owner can verify the ownership of the target data.   
\par Existing IPP techniques are limited to confirming model ownership but cannot establish a clear association between the ML model and training data. 
To bridge this gap, Liu \textit{et al.} \cite{liu2022your} introduced MeFA, which imported membership inference techniques to create fingerprint data for IP verification. The main concept is to select training samples with the greatest influence on the model’s prediction behavior as member fingerprints. Tian \textit{et al.} \cite{tian2023knowledge} believed that the knowledge transferred from a training dataset to a DL model could be uniquely represented by the model’s DB.  Therefore, the authors designed a novel generation method that utilizes geometric consistency to find the samples supporting the DB, which can serve as the proxy for the knowledge representation. Specifically, geometric perspective refers to the direction consistency between the normal vector of the DB and the vector from the training sample to its corresponding boundary samples.

\par The above two prediction behavior-based methods have two significant advantages: (i) They require no modification to the target data or the training process; (ii) It is a model-agnostic framework that can be applied to any type of DL model and does not necessitate prior knowledge of the suspect model’s type, structure, parameters, or training process.

\subsection{Dataset Authorization Control}
\par As a proactive IPP method, the data owner can disrupt illegal use by adding imperceptible or reversible perturbations.
\par The work \cite{xue2023dataset} first generates adversarial examples by perturbing the images in feature space. The clean image was then hidden in the matching perturbed image by employing the modified reversible image transformation technique to produce protected images. Following the transformation process, a secret key will be created and safely kept. Hence, for unauthorized users without the secret key, the protected dataset is directly used to train the unauthorized model, which will make the trained model have a very poor inference accuracy.

\par Recent efforts have been made to finetune open-source text-to-image DMs with additional samples from specific artists to generate AI art that mimics the specific artistic style of the artist. Shan \textit{et al.} \cite{shan2023glaze} argued that this behavior essentially infringes on the copyrights of the artists, as the artistic style is also an IP form that needs to be protected. Thus, Glaze is devised to allow artists to add imperceptible perturbations to their artworks from style mimicry before sharing them online. This operation, on the one hand, does not affect the visual effect of the artworks, while on the other hand, it can mislead generative models to produce artworks that differ from the style of the target artist.

\section{Distributed Intellectual Property Protection}
\label{5}

\subsection{Overview}
\par As the demand for processing training data has outstripped the growth in computational capabilities of individual machinery \cite{zhou2020privacy}, there is a need to transition from centralized ML to distributed ML (DML). In addition, DML is primarily motivated by privacy-preserving incentives, ensuring that local private data remains on the device and is not exposed to external entities. Various DML schemes exist, including federated learning (FL), split learning (SL), peer-to-peer learning, and private aggregation of teacher ensembles (PATE). Although data is not explicitly exchanged in the DML, the training process requires the sharing of information about participants’ models. This makes it trivial for a malicious actor to steal or distribute the individual model without authorization.

\par However, the majority of IPP methods now in use are created within centralized learning frameworks and are unsuitable for DML due to differences in data access privileges. Although DML has its advantages, it also brings forth the issue of preserving ownership. Since the FL situation is the primary focus of existing IPP techniques, we shall analyze the challenges presented by DML by taking FL as an example:

\begin{itemize}
\item The fact that the server cannot access the client's dataset and only learns about the target task poses a significant challenge to IPP methods that require access to the raw training data to construct query samples.

\item In centralized learning, it is only necessary to perform a single injection of IP identifiers. However, FL involves multiple rounds of communication, so IPP injection from the beginning to the end is required.

\item Multiple clients, some of whom may be malicious, may join forces to launch the collusion attack, which enables them to gain a faster and more comprehensive understanding of the model by sharing information they have obtained and thus jointly bypassing or identifying IPP mechanisms.

\item Conflicts between the watermarks might easily arise when multiple clients are permitted to inject watermarks. Watermark collisions can also easily happen, even when distinct trigger sets are intended for different clients.

\item How to ensure that embedded watermarks and fingerprints are compatible with various privacy-preserving approaches (e.g., differential privacy, secure aggregation) and client-side selection strategies is also an ongoing question.

\item Both communication and computation overhead are concerns considering the FL clients who are often resource-restricted.
\end{itemize}

\subsection{Federated Model IPP} 
\begin{figure*}[htbp]
	\centering
	\includegraphics[width=.75\textwidth]{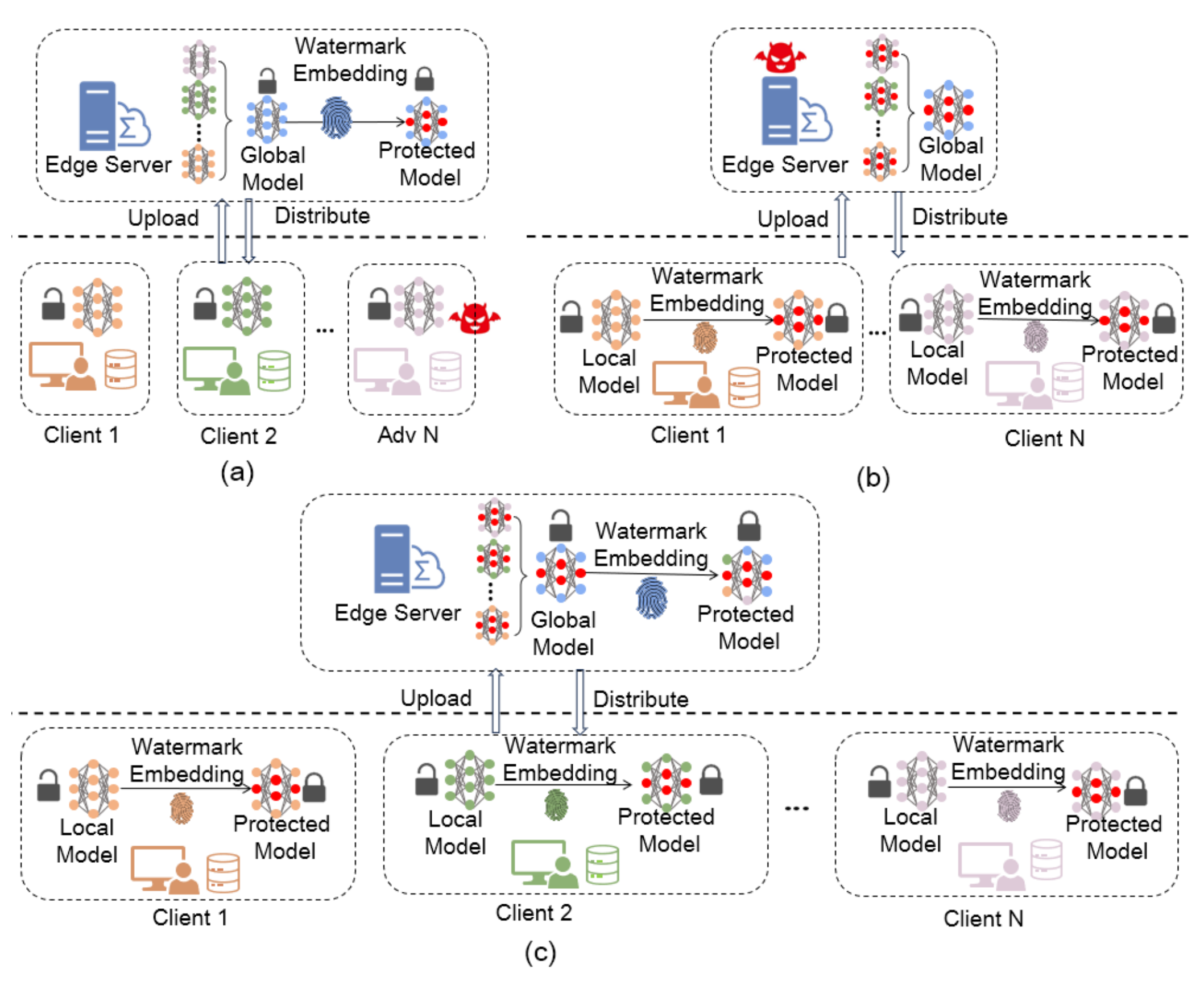}
 \vspace{-0.1in}
    \caption{An example of each possible scenario of watermarking in FL with one server and three clients. (a) Server-side model IPP. (b) Client-side model IPP. (c) Collaborative federated model IPP.}
    \label{fig:FL mode}
 \vspace{-0.1in}
\end{figure*}

The first question in the client-server FL framework is which party can add the ownership credential to the model and needs to be protected. The parties responsible for injecting the IP identifiers are the IP owners. We distinguish the following three FL IPP scenarios: server-side, client-side, and collaborative federated model IPP (as shown in Fig. \ref{fig:FL mode}). The server-side IPPs assume that the main adversaries are malicious clients and protect the server IP effort. Meanwhile, client-side IPPs target the semi-honest server and require multiple clients to inject their own watermarks. The collaborative IPPs, on the other hand, are suitable for scenarios with a high risk of privacy leakage and assume that both the clients and aggregator may not be trustworthy. The summary of federated model IPP approaches is presented in Table \ref{table:FED comparison}.

\begin{table*}[]
\centering
\caption{Summary of federated model IPP approaches.}
\label{table:FED comparison}
\scalebox{0.7}{
\begin{threeparttable}  
\begin{tabular}{|c|c|c|cccc|c|c|}
\hline
\multicolumn{1}{|c|}{\multirow{2}{*}{\textbf{Category}}} & \multicolumn{1}{c|}{\multirow{2}{*}{\textbf{Paper}}} & \multirow{2}{*}{\textbf{Model Access}}                            & \multicolumn{4}{c|}{\textbf{Security Tools Compatibility}}                                                                                                                               & \multicolumn{1}{c|}{\multirow{2}{*}{\textbf{\begin{tabular}[c]{@{}c@{}}IP Identifer\\ Conflicts\end{tabular}}}} & \multicolumn{1}{c|}{\multirow{2}{*}{\textbf{Robustness}}} \\ \cline{4-7}
\multicolumn{1}{|c|}{}                                   & \multicolumn{1}{c|}{}                                &                                                                   & \multicolumn{1}{c|}{DP\textsuperscript{1}}
                        & \multicolumn{1}{c|}{HE\textsuperscript{2}}                           & \multicolumn{1}{c|}{MPC\textsuperscript{3}}                          & \multicolumn{1}{c|}{CS\textsuperscript{4}}                          & \multicolumn{1}{c|}{}                                                                                           & \multicolumn{1}{c|}{}                                     \\ \hline
\multirow{4}{*}{Server-side}                             & \cite{tekgul2021waffle}                              & Black box                                                         & \multicolumn{1}{c|}{Medium} & \multicolumn{1}{c|}{Poor} & \multicolumn{1}{c|}{Medium} & \multicolumn{1}{c|}{Medium} & \multicolumn{1}{c|}{Medium}                                                                                    & \CIRCLE\LEFTcircle\Circle\Circle                              \\ \cline{2-9} & \cite{shao2024fedtracker}                            & \begin{tabular}[c]{@{}c@{}}White \& \\ Black box\end{tabular} & \multicolumn{1}{c|}{Strong} & \multicolumn{1}{c|}{Poor} & \multicolumn{1}{c|}{Medium} & \multicolumn{1}{c|}{Medium} & \multicolumn{1}{c|}{Poor}                                                                                    & \CIRCLE\CIRCLE\Circle\Circle                              \\ \cline{2-9} & \cite{fan2023fate}                                   & \begin{tabular}[c]{@{}c@{}}White \& \\Black box\end{tabular} & \multicolumn{1}{c|}{Strong} & \multicolumn{1}{c|}{Poor} & \multicolumn{1}{c|}{Strong} & \multicolumn{1}{c|}{Medium} & \multicolumn{1}{c|}{Poor}                                                                                  & \CIRCLE\CIRCLE\CIRCLE\Circle                              \\ \cline{2-9}  & \cite{chen2024fedmct}                                & \begin{tabular}[c]{@{}c@{}}White \& \\Black box\end{tabular} & \multicolumn{1}{c|}{Medium} & \multicolumn{1}{c|}{Poor} & \multicolumn{1}{c|}{Medium} & \multicolumn{1}{c|}{Medium} & \multicolumn{1}{c|}{Medium}                                                                                     & \CIRCLE\CIRCLE\Circle\Circle                              \\ \hline
\multirow{7}{*}{Client-side}                             & \cite{li2022fedipr}                                  & \begin{tabular}[c]{@{}c@{}}White \& \\Black box\end{tabular} &\multicolumn{1}{c|}{Strong} & \multicolumn{1}{c|}{Medium} & \multicolumn{1}{c|}{Medium} & \multicolumn{1}{c|}{Strong} & \multicolumn{1}{c|}{Medium}                                                                                    & \CIRCLE\CIRCLE\LEFTcircle\Circle                              \\ \cline{2-9} & \cite{yang2023watermarking}                          & Black box                                                        &\multicolumn{1}{c|}{Medium} & \multicolumn{1}{c|}{Strong} & \multicolumn{1}{c|}{Medium} & \multicolumn{1}{c|}{Medium} & \multicolumn{1}{c|}{Strong}                                                                                    & \CIRCLE\CIRCLE\LEFTcircle\Circle                              \\ \cline{2-9}  & \cite{nie2024persistverify}                          & Black box                                                         &\multicolumn{1}{c|}{Poor} & \multicolumn{1}{c|}{Medium} & \multicolumn{1}{c|}{Medium} & \multicolumn{1}{c|}{Medium} & \multicolumn{1}{c|}{Strong}                                                                                      & \CIRCLE\CIRCLE\CIRCLE\Circle                              \\ \cline{2-9} & \cite{liu2021secure}                                 & Black box                                                         &\multicolumn{1}{c|}{Medium} & \multicolumn{1}{c|}{Strong} & \multicolumn{1}{c|}{Medium} & \multicolumn{1}{c|}{Medium} & \multicolumn{1}{c|}{Strong}                                                                                     & \CIRCLE\CIRCLE\Circle\Circle                              \\ \cline{2-9}  & \cite{liang2023fedcip}                               & White box                                                         &\multicolumn{1}{c|}{Medium} & \multicolumn{1}{c|}{Poor} & \multicolumn{1}{c|}{Medium} & \multicolumn{1}{c|}{Medium} & \multicolumn{1}{c|}{Medium}                                                                                     & \CIRCLE\CIRCLE\Circle\Circle                              \\ \cline{2-9} & \cite{yang2023fedzkp}                                & White box                                                        &\multicolumn{1}{c|}{Medium} & \multicolumn{1}{c|}{Strong} & \multicolumn{1}{c|}{Strong} & \multicolumn{1}{c|}{Medium} & \multicolumn{1}{c|}{Medium}                                                                                    & \CIRCLE\CIRCLE\LEFTcircle\Circle                              \\ \cline{2-9}  & \cite{yu2023leaked}                                  & Black box                                                         &\multicolumn{1}{c|}{Medium} & \multicolumn{1}{c|}{Strong} & \multicolumn{1}{c|}{Strong} & \multicolumn{1}{c|}{Medium} & \multicolumn{1}{c|}{Medium}                                                                                     & \CIRCLE\CIRCLE\CIRCLE\Circle                              \\ \hline
\multirow{3}{*}{Collaborative}                           & \cite{chen2023fedright}                              & Black box                                                         &\multicolumn{1}{c|}{Medium} & \multicolumn{1}{c|}{Poor} & \multicolumn{1}{c|}{Medium} & \multicolumn{1}{c|}{Medium} & \multicolumn{1}{c|}{Strong}                                                                                    & \CIRCLE\CIRCLE\Circle\Circle                              \\ \cline{2-9}  & \cite{li2021towards}                                 & \begin{tabular}[c]{@{}c@{}}White \& \\Black box\end{tabular} &\multicolumn{1}{c|}{Medium} & \multicolumn{1}{c|}{Poor} & \multicolumn{1}{c|}{Medium} & \multicolumn{1}{c|}{Medium} & \multicolumn{1}{c|}{Medium}                                                                                  & \CIRCLE\CIRCLE\LEFTcircle\Circle                              \\ \cline{2-9}  & \cite{luo2024copyright}                              & White box                                                         &\multicolumn{1}{c|}{Medium} & \multicolumn{1}{c|}{Poor} & \multicolumn{1}{c|}{Poor} & \multicolumn{1}{c|}{Medium} & \multicolumn{1}{c|}{Medium}                                                                                   & \CIRCLE\CIRCLE\CIRCLE\Circle                              \\ \hline
\end{tabular}
\begin{tablenotes}
\centering
\item 
    \quad(1) DP -- Differential Privacy \quad(2) HE -- Homomorphic Encryption  \quad(3) MPC -- Multi-Party Computation \quad(4) CS -- Client Selection 
    \item  Poor, Medium, and Strong denote the compatibility of the federated IPP approaches with security protocols (Strong$>$Medium$>$Poor).
    \item The more $\CIRCLE$ indicates that the approach proposed in the work is more capable in this aspect.  
\end{tablenotes}
\end{threeparttable}
}
\end{table*}

\par \textit{1) Server-side federated model IPP.} The server, as the IP owner, is responsible for creating a fingerprint or watermark for the federated global model.
\par WAFFLE \cite{tekgul2021waffle} is the first solution for addressing ownership problems in client-server FL. It embeds a backdoor-based watermark by re-training the global model during each aggregation round. The trigger set is produced in a data-independent way by inserting random but class-consistent patterns on the backdoor sample’s background. 
However, embedding triggers outside the FL task distribution will result in a significant decrease in model performance. To address this problem, FedTracker \cite{shao2024fedtracker} was proposed to provide traceable ownership in FL scenarios. It employs a bi-level protection strategy, in which the server is responsible for embedding a WAFFLE watermark and generating a random secret matrix and a unique fingerprint for each client. During the verification phase, the server infers traitors by comparing the fingerprint similarity.

\par \textit{2) Client-side federated model IPP.} More than one client adds ownership certificates to their local models to inject watermarks or fingerprints into the global model.
\par FedIPR \cite{li2022fedipr} is the first work that enables all clients to incorporate their unique watermarks into the global model. It offers both black box and white box ways, with the embedding process completed during the client’s local optimization. white box watermarks are embedded by adding a regularization term to the primary task’s loss function, while black box watermarks are implemented by introducing backdoor triggers as additional inputs. Building on ~\cite{li2022fedipr}, Yang \textit{et al.} \cite{yang2023watermarking} designed a black box watermarking scheme based on the client-side backdoor, where a pre-designed trigger set is embedded into the FL model via gradient-enhanced embedding method. The main contribution is the proposed trigger set construction pattern, which ensures that the watermark cannot be forged. Instead of using the classical Gaussian noise-based trigger set, it utilizes a permutation-based secret key and noise-based pattern to construct the backdoor images. Compared to the method in \cite{li2022fedipr}, this new method offers enhanced defense performance and robustness against various watermark removal and ambiguity attacks. 

\par Existing IPP solutions cannot address the issue that new theft behaviors by undetected malicious clients may occur in each communication round within the FL framework. To this end, FedCIP \cite{liang2023fedcip} developed the concept of periodic watermarking. Specifically, communication rounds are divided into different periods, with distinct watermarks embedded in each period. Notably, within each period, the same watermark is embedded into the models of all currently selected clients. The core idea for tracking traitors is that if a client skips training during a particular period, its local model will lack this period watermark. Each federated watermark is a unique period identifier; by analyzing the watermarks, it is possible to determine which period the leaked model originated from, thereby identifying the potential set of traitors.

\par To precisely identify the traitor of a leaked model, a straightforward approach is to inject different watermarks for different clients. However, this operation also brings a series of negative effects. For example, the increase in forged knowledge significantly raises the probability of watermark collisions among clients, which in turn reduces the model’s utility and robustness. For this reason, Yu \textit{et al.} \cite{yu2023leaked} proposed the decodable unique watermark (DUW) to resolve the watermark collision pitfall, which comprised two steps. In the first step, a unique key corresponding to each client ID is designed as a one-hot binary string to distinguish the clients. For each client, by providing the unique key and OOD data to a pre-trained encoder, the key can be embedded into the backdoor sample as a sample-specific trigger. In the second step, to avoid conflicts from label repetition, it is suggested to project the output dimension of the original model to a higher dimension, allowing each client to have a unique target label.

\par \textit{3) Collaborative federated model IPP.} The server and the clients collaborate to watermark the global model together.
\par FedRight \cite{chen2023fedright} is the first to introduce model fingerprints for FL IPP. The generation of these model fingerprints primarily involves two components: selected adversarial samples and extracted model-specific features. The former serves merely as carriers, while the latter is the extracted model-specific features, reflecting the model’s unique information. It is noted that the selected adversarial samples are independent of the federated task. This precaution prevents attackers from launching adversarial retraining attacks, which could compromise the effectiveness of ownership verification. Merkle-Sign, proposed by Li \textit{et al.} \cite{li2021towards}, is a public authentication protocol based on the Merkle tree. In this framework, the server inserts identity information (i.e., keys) into the global model during each aggregation phase. Meanwhile, the server uploads the key tuple and authentication function tuple (generated by the watermark embedding function) into the Merkle tree over time. At last, the server also embeds all the clients’ keys into the final model and updates the Merkle tree accordingly. As the client’s identity information is recorded, the server can track the traitor based on the extracted watermarking.

\par The current IPP framework in the federation typically works well in scenarios where malicious clients act alone. Still, it cannot accurately identify malicious clients when the model is subjected to collusion attacks. To fill this gap, Luo \textit{et al.} \cite{luo2024copyright} proposed a new anti-collusion attack watermark protection scheme, named FedFP. By adopting the anti-collusion coding theory, unique watermark information is designed for each client, thereby effectively detecting colluders. Additionally, it utilizes a specific regularization loss function to embed watermark information and combines skip connections to embed the watermark information into each batch normalization layer.

\section{Attacks on Deep IPP}
\label{7}
\subsection{Overview of Threats}
\par  The IPP for model and data has taken on new significance in the digital age. But these so-called defense mechanisms might open up new attack vectors, leaving the protected models even more vulnerable to infringement than the unprotected models. For instance, watermarks may significantly alter the distribution of model weights, and adversaries may be able to identify and manipulate the watermarks. As shown in Table \ref{table:attacks}, a large number of methods have been proposed to invalidate the IP identifiers. We categorize the different types of attacks against DL model IPP methods into the following two levels (from weak to strong): 1) IP detection and evasion; and 2) IP removal. 

\begin{table*}[]
\centering
\caption{Summary of attacks on deep IPPs.}
\label{table:attacks}
\scalebox{0.7}{
\begin{tabular}{|cc|c|c|l|l|}
\hline
\multicolumn{2}{|c|}{\textbf{Category}}                                                                                                                                                                                                                                                                                                           & \textbf{\begin{tabular}[c]{@{}c@{}}Attack \\ Level\end{tabular}} & \textbf{\begin{tabular}[c]{@{}c@{}}Model \\ Access\end{tabular}} & \multicolumn{1}{c|}{\textbf{Purposes}}                                                                                                              & \multicolumn{1}{c|}{\textbf{Limitations}}                                                                                 \\ \hline
\multicolumn{1}{|c|}{\multirow{4}{*}{\begin{tabular}[c]{@{}c@{}}IP Detection \\ \& Evasion\end{tabular}}} & \begin{tabular}[c]{@{}c@{}}Model property detection\\ \cite{wang2021riga,wang2019attacks,hitaj2019evasion}\end{tabular}                                                                                            & \multirow{4}{*}{Level 1}                                         & White box                                                         & \multirow{2}{*}{\begin{tabular}[c]{@{}l@{}}Exploit these anomalous attributes to\\ detect the embedded IP identifiers\\ and evade ownership verification.\end{tabular}} & Have the highest-level model access.                                                                          \\ \cline{2-2} \cline{4-4} \cline{6-6} 
\multicolumn{1}{|c|}{}                                                                                    & \begin{tabular}[c]{@{}c@{}}Model behavior detection\\ \cite{wang2022rethinking,sun2021detect,xu2017feature,namba2019robust}\end{tabular}                                                                             &                                                                  & Black box                                                         &                                                                                                                                                     & Know the crafting means of key triggers.                                                                                  \\ \cline{2-2} \cline{4-6} 
\multicolumn{1}{|c|}{}                                                                                    & \begin{tabular}[c]{@{}c@{}}IP ambiguity \\ \cite{fan2019rethinking,uchida2017embedding,fan2021deepipr,chen2023effective}\end{tabular}                                                                              &                                                                  & \multirow{2}{*}{White box}                                        & \multirow{2}{*}{\begin{tabular}[c]{@{}l@{}}Generate substitute IP identifiers\\ to confuse ownership.\end{tabular}}                                 & \multirow{2}{*}{\begin{tabular}[c]{@{}l@{}}Easily encounter significant performance \\ degradation.\end{tabular}}         \\ \cline{2-2}
\multicolumn{1}{|c|}{}                                                                                    & \begin{tabular}[c]{@{}c@{}}Collusion attack \cite{chen2019deepmarks}\end{tabular}                                                                                                                                                &                                                                  &                                                                   &                                                                                                                                                     &                                                                                                                           \\ \hline
\multicolumn{1}{|c|}{\multirow{2}{*}{IP Removal}}                                                         & \begin{tabular}[c]{@{}c@{}}Model modification\\ \cite{chen2019leveraging,chen2021refit,guo2020fine,20240115514}\end{tabular}                                                      & \multirow{2}{*}{Level 2}                                         & White box                                                         & \multirow{2}{*}{\begin{tabular}[c]{@{}l@{}}Remove IP identifiers and create\\ pirated models possessing the same \\ performance.\end{tabular}}             & \multirow{2}{*}{\begin{tabular}[c]{@{}l@{}}Collect a large amount of data for fine-tuning \\ or retraining.\end{tabular}} \\ \cline{2-2} \cline{4-4}
\multicolumn{1}{|c|}{}                                                                                    & \begin{tabular}[c]{@{}c@{}}Model extraction\\ \cite{lin2023quda,chen2021stealing,xue2024removing,zong2024ipremover,shen2022model,patwari2022dnn}\end{tabular} &                                                                  & Black box                                                         &                                                                                                                                                     &                                                                                                                           \\ \hline
\end{tabular}}
\end{table*}

\subsection{Level 1: IP Detection \& Evasion}
\par \textit{1) Threat model:}
\par $\bullet$ \textbf{Adversary’s goal:} When embedding white box watermarks, the model’s parameters or neuron states are often modified to some degree. In the case of black box watermarks, key samples typically contain specific trigger patterns, and the target labels are often manipulated. Therefore, attackers aim to exploit these anomalous attributes to detect the embedded watermarks or fingerprints, understand their operational model, and ultimately evade ownership verification.

\par $\bullet$ \textbf{Assumptions:} Attackers possess full access, including both parameters and structures, to multiple models that have been trained on datasets from the same domain. Besides, the attacker possesses a local shadow dataset, which can be used as a control group to analyze the abnormal behavior of query inputs.

\par \textit{2) Attack methods:}
\par $\bullet$ \textbf{Model property detection.} Attackers can analyze the presence of watermarks by examining the distribution of model weights, intermediate outputs, and neural network states. These schemes require attackers to have the highest-level access to the model.

\par RIGA \cite{wang2021riga} reported that existing white box watermarking algorithms change the weight distribution of the target model, allowing attackers to detect watermarks simply by visual inspection. Further, Wang \textit{et al.} \cite{wang2019attacks} proposed an attack method that can derive the watermark embedding length and overwrite the original watermark. In work \cite{hitaj2019evasion}, the attacker collected or stole multiple trained DL models that perform the same task to build an ensemble of models. Given a new input instance, the attacker observed the hidden layer activations of these reference models. For models that are not watermarked, the predicted class of the input instance can be considered a random event. Otherwise, the watermark is deemed to be present. However, these methods require sufficient samples of the weight distribution and demand high privileges from the attacker.

\par $\bullet$ \textbf{Model behavior detection.} Attackers can detect and remove watermarks by reversing data triggers or injecting destructive perturbations. IP detection attacks against input samples are typically conducted under the condition that the means of sample generation are already known.
\par AdVNP \cite{wang2022rethinking} is an input preprocessing attack by injecting naturalness-aware perturbations to invalidate the watermark. The key insight is that watermarks with visible/invisible triggers that are learned from divergent distributions are more susceptible to naturalness-aware perturbations of similar magnitudes. The work \cite{sun2021detect} employs GANs to generate simulated perturbations to enable the watermark model to misclassify perturbed images with maximum probability. The effectiveness of the aforementioned two attacks is relatively weak, and they require a significant amount of computational resources. Xu \textit{et al.} \cite{xu2017feature} proposed a feature-squeezing framework for detecting adversarial samples. This framework coalesces samples corresponding to various feature vectors in the original space into a single sample, thereby reducing the space that needs to be searched during watermark detection. If the difference in predictions between the feature-squeezed input and the original input exceeds a threshold level, the input is identified as adversarial.
\par Targeted at pattern-based backdoor watermarking, Namba \textit{et al.} \cite{namba2019robust} presented query modification, a novel technique for evading watermark detection. Naturally, an autoencoder can subtract or distort specific kinds of key samples that are placed on the picture. The Jensen-Shannon (JS) divergence between the samples before and after reconstruction is first noticed by feeding the query into the autoencoder. Next, a thorough analysis is conducted of the differences between the projected label distributions before and after the autoencoder is applied. These two observed measures will exhibit a considerable difference if the query is a key sample.

\par $\bullet$ \textbf{IP ambiguity.} The attacker aims to lower the confidence of verification results by forging an additional IP identifier to overwrite or invalidate the original one.
\par The pioneering work was conducted by Fan \textit{et al.} in work \cite{fan2019rethinking}, which highlighted the vulnerability of Uchida’s approach \cite{uchida2017embedding} under ambiguity attacks. It was explained that traditional watermarking methods can be forged as long as the model’s performance is independent of the signature \cite{fan2021deepipr}. Based on this proposition, Fan and his team designed a passport layer through which the model’s functionality is controlled by a signature called a passport. However, this solution encountered significant performance degradation when batch normalization layers were present. Further, Chen \textit{et al.} \cite{chen2023effective} proposed a novel and effective ambiguity attack method that successfully forged multiple valid passports using a small training dataset. This was accomplished by inserting a specially designed attachment block ahead of the passport parameters.

\par $\bullet$ \textbf{Collusion attack.}  It is an active attack. The group of authorized users with identical host DL models and distinct IP identifiers may engage in collusion attacks, aiming to develop a model with equivalent functionality or collude with unauthorized users to grant authorized access to multiple unauthorized users.
\par Chen \textit{et al.} \cite{chen2019deepmarks} proposed a typical and cost-effective FP collusion attack known as the FP averaging attack. Colluders are aware of the location of the watermark embedding layer, and the colluders can perform element-wise averaging on their weights and present the generated average as a response to model owners' queries.

\subsection{Level 2: IP Removal}
\par \textit{1) Threat model:}

\par $\bullet$ \textbf{Adversary’s goal:} Malicious attackers aim to remove built-in IP identifiers and create pirated models without watermarks or fingerprints while possessing the same performance as the pirated models.

\par $\bullet$ \textbf{Assumptions:} Adversaries can gain access with either white box or black box privileges. With white box access, adversaries can adjust the model weights or structure, such as through model fine-tuning, fine pruning, regularization, and neural cleansing. With black box access, adversaries learn a pirated model from the protected model's predictions or properties, but they require sufficient in-distribution data to support this learning.

\par \textit{2) Attack methods:}
\par $\bullet$ \textbf{Model modification.} Recent studies suggested using either fine-tuning \cite{chen2019leveraging,chen2021refit} or training regularization \cite{guo2020fine,20240115514} to remove watermarks. 
\par The first method can reduce the quantity of labeled training data required for successful watermark removal by using the additional unlabeled data to fine-tune the watermarked model \cite{chen2019leveraging}. But in reality, gathering this much more unlabeled data presents a significant challenge to enemies. REFIT \cite{chen2021refit} demonstrated that while using a high learning rate during fine-tuning can successfully remove the watermark, it significantly compromises test accuracy. Although elastic weight consolidation (EWC) was employed by REFIT to mitigate model performance degradation, the effectiveness of this method heavily relies on a carefully designed learning rate schedule, which poses challenges for broader generalization.
\par However, works \cite{chen2019leveraging,chen2021refit} need to know the type of the watermark to adjust the fine-tuning weights. The attacks become less plausible or realistic as a result of these strong presumptions. Instead of just fine-tuning the watermarked models, Guo \textit{et al.} \cite{guo2020fine} combined imperceptible pattern embedding and spatial-level transformations to create a straightforward yet potent transformation algorithm that can effectively and blindly destroy the memorization of watermarked models to the watermark samples. Compared to previous works, this method requires much less information regarding the watermarking technique and resources. Pegoraro \textit{et al.} \cite{20240115514} focused on breaking the white box DL model watermarking scheme, and in turn proposed DeepEclipse, which first identified possible watermarking layers using frequency analysis and then applied more substantial model modifications to make the identified layers noisy.

\par $\bullet$ \textbf{Model extraction.} Model extraction learns model copies from model predictions or attributes, such as query-based steganography attacks and side-channel attacks. Typically, model extraction has stronger performance in IP removal, but requires larger datasets.
\par The existing extraction methods, while having been proven to achieve high cloning accuracy, come at a high cost of extraction. Defenders can easily defend by limiting the number of queries. Lin \textit{et al.} \cite{lin2023quda} proposed QUDA, a novel queue-restricted data-free model extraction attack. It combines a GAN pre-trained on a public unrelated dataset to provide weak image priors and deep reinforcement learning techniques to improve the efficiency of the query generation strategy. It can be applied to scenarios without data and with restrictions on the number of queries. Due to the high complexity and the limited observable information, model extraction attacks against supervised DL models cannot be applied to Deep reinforcement learning (DRL) scenarios. To this end, Chen \textit{et al.} \cite{chen2021stealing} presents the first model extraction attack in the DRL setting, which enables attackers to precisely recover a black box DRL model solely from interactions with the environment.
\par For image processing networks, Xue \textit{et al.} \cite{xue2024removing} introduced a watermark extraction approach that reconstructs the model output image using an asymmetric UNet and further uses the reconstructed version for model retraining. In particular, an attention module for reference subspace is suggested and incorporated into the asymmetric UNet. This module eliminates the watermark by projecting the target model's output into a subspace of the reference image. From the attacker’s perspective, a successful IP removal attack should simultaneously defeat fingerprinting and watermarking. However, to the best of our knowledge, work on such attacks is scarce. The IP removal attack proposed by Zong \textit{et al.} \cite{zong2024ipremover} filled this gap. Little attention has been paid to models trained with graph data (i.e., Graph neural networks), most of them focus on the models trained with images and texts. Shen \textit{et al.} \cite{shen2022model} were the first to propose a model stealing attack for inductive graph neural networks (GNN) to fill this gap. Many model owners treat the unique structure of the model as a fingerprint. In response to this type of scheme, Patwari \textit{et al.} \cite{patwari2022dnn} infer the model structure by observing CPU/GPU-related information during the model’s operation.

\section{Challenges and Prospects}
\label{8}
\par Here, we draw attention to the challenges associated with deep IPPs and prospect promising future directions that may act as a guide for innovative research. We prioritize challenges and directions with significant practical impact, introducing them in order of importance.

\subsection{Usable Unified Verification Metrics}
\par Unified evaluation metrics are fundamental for the practical implementation of deep IPPs. However, we perceive that the existing verification metrics have the following shortcomings: (1) The existing evaluation standards are mostly targeted at reactive schemes, lacking assessment of proactive IPP represented by proactive authorization control. (2) Evaluation is mostly focused on performance metrics (such as fidelity, capacity, etc.), while research on anti-attack metrics (such as security, non-removability, etc) is insufficient. Therefore, designing a safe, effective, comprehensive verification metric that can serve as a unified standard is worth careful consideration. 

\subsection{Model and Dataset Intelligence IPP}
\par In the current DL field, deep IPP should be extended from the model level to the dataset level. Specifically, training datasets, especially those that have been carefully collected, cleaned, and annotated, are also an important part of a participant’s core competitiveness. In addition, the output data of DL models, such as generated images, voice, and text, also have certain artistic value and market potential. For example, paintings created by AI models can, in some cases, rival or even surpass the works of human artists, and their need for IPP is equally urgent \cite{shan2023glaze}. Therefore, how to effectively manage and protect the IP of these datasets to prevent unauthorized access, copying, or use has become an urgent problem that needs to be solved.

\subsection{Beyond Classification Tasks}
\par Current IPPs focus primarily on classification tasks. However, few works focus on more complex and increasingly prevalent types of tasks, such as object detection, image generation and segmentation, speech recognition, and text generation that are built on vast datasets and complex architectures. These models not only push the limits of AI, but they also create new technical challenges. For example, they make it easier to process data across multiple modalities (like text and images together). Implementing IPP in these advanced models faces multiple challenges. Therefore, future research on deep IPP should broaden its perspective, not only focusing on basic classification tasks but also delving into protection strategies for complex task scenarios and large foundation models.

\subsection{Theoretical Analysis and Proof}
\par The current deep IPP methods still lack a rigorous theoretical analysis, which significantly reduces the rigor and authority of these methods in practical applications. Their defense effectiveness largely depends on the assumptions about the threat model and simulation results, and black box access further increases the difficulty of providing theoretical guarantees \cite{hu2023unbiased}.
\par According to our summary, several promising sub-issues are particularly worth studying in the future: (1) The core motivation of fingerprint recognition methods is to select or produce samples located near the decision boundary to uniquely represent a DL model. How can we prove that the distance from the fingerprint triggers to the model decision boundary is bounded and within the range that can ensure effectiveness? (2) How many bits of watermark can a defender embed in the host model without affecting the model’s ability to perform its original task? (3) How to prove the robustness of such methods that embed watermarks into the model’s redundant space when against attacks such as adversarial fine-tuning, model pruning, etc.? (4) In a distributed learning scenario, how many users can the system tolerate embed their own watermarks or fingerprint information without conflicts? Studying these issues will provide important support for the theoretical foundation and practical application of deep IPP methods.

\subsection{Effectiveness and Efficiency}
\par To date, whether in centralized or distributed scenarios, the construction of IP identifiers typically involves extensive computation and optimization on models or samples. Currently, research on distributed learning is still relatively limited. First, the research on distributed IPP is mainly focused on the FL setting. However, other distributed settings, such as SL, collaborative learning, and peer-to-peer learning, also require operations such as embedding watermarks, extracting fingerprints, and access control. Second, existing FL IPP methods are difficult to ensure efficiency, integrity, and robustness simultaneously. In addition, the exchange of intermediate information is required during the distributed IPP process, leading to excessive consumption of time and resources. Therefore, developing IPP methods that can balance effectiveness, efficiency, and fidelity across diverse distributed learning is crucial.

\subsection{Multi-Modal LLM Intelligence IPP}
Multi-modal LLMs (MM-LLMs) are more vulnerable to model stealing attacks \cite{liu2024survey}, as they require significantly more training data and resources to achieve high performance compared to single-modal LLMs (SM-LLMs). Currently, research on watermarking and fingerprinting algorithms for MM-LLMs lags significantly behind SM-LLMs due to the following challenges: (1) Directly transferring single-modal IPPs to MM-LLMs may significantly reduce the protected model functionality. For instance, in vision-language pre-trained LLMs, which provide semantic-aligned visual and language embeddings, the watermark must maintain the transformation correlation between the two modalities. However, single-modal watermarks may disrupt this embedding correlation, leading to a decline in performance \cite{tang2023watermarking}. (2) The complexity of MM-LLMs and inherent conflicts between multiple design objectives make it challenging to achieve trade-offs between the IPPs' covertness, robustness, and fidelity. Thus, the development of deep IPP algorithms applicable to MM-LLMs is an essential future research direction.

\section{Conclusion}
\label{9}
Well-trained models and pre-constructed datasets are valuable assets but are vulnerable to risks such as illegal reproduction, theft, redistribution, and misuse. Acknowledging these challenges, we have provided a comprehensive overview of deep IPP, addressing the critical need for IPP, its criteria, taxonomy, distributed settings, applications, threats, and prospects. Specifically, we systematically differentiated between evaluation metrics that are common and those specific to reactive and proactive schemes. For IPP on both models and datasets, we offered a detailed sub-categorization based on the diverse strategies employed. This survey also underscored the challenges and solutions related to deep IPP in distributed deep learning settings. We have also elaborated on attacks targeting IPP mechanisms. Lastly, we outlined promising future research directions.

\section{Acknowledgment} \label{Acknowledgment}

This work is supported by National Natural Science Foundation of China (62072239, 62372236, and 62402223), Open Foundation of the State Key Laboratory of Integrated Services Networks (ISN24-15), the Postdoctoral Fellowship Program of CPSF (GZB20240982), Qing Lan Project of Jiangsu Provenince, and Jiangsu Funding Program for Excellent Postdoctoral Talent.

\bibliographystyle{acm_reference_format}
\bibliography{sample_base}

\end{document}